\documentclass[aps,twocolumn,prb,amsmath,amssymb,superscriptaddress,floats]{revtex4}

\usepackage{amsmath}	
\usepackage{gensymb}
\usepackage{hyperref}
\hypersetup{colorlinks=true}
\usepackage{upgreek}
\usepackage{graphics}
\usepackage{hyperref}
\usepackage{epsfig}
\usepackage{color}
\usepackage{bm}
\usepackage{float}
\usepackage{ulem}               
\usepackage{dsfont}             
\usepackage{lettrine}           

\usepackage{CJK}
\usepackage{url}
\usepackage{multirow}

\graphicspath{{./}{./Figures/}}

\usepackage{titlesec}
\titleformat*{\section}{\normalfont\normalsize\bfseries\sffamily}
\titleformat{\subsection}[runin]
  {\normalfont\normalsize\bfseries}{\thesubsection}{1em}{}
\titlespacing*{\section}
  {0pt}{5.5ex plus 1ex minus .2ex}{0.3ex plus .2ex}

\begin{document}

\title{
Nonsymmorphic bosonization in one-dimensional generalized Kitaev spin-1/2 models
}

\author{Wang Yang}
\thanks{These two authors contributed equally to this work.}
\affiliation{Department of Physics and Astronomy and Stewart Blusson Quantum Matter Institute,
University of British Columbia, Vancouver, B.C., Canada, V6T 1Z1}

\author{Chao Xu}
\thanks{These two authors contributed equally to this work.}
\affiliation{Kavli Institute for Theoretical Sciences, University of Chinese Academy of Sciences, Beijing 100190, China}

\author{Shenglong Xu}
\affiliation{Department of Physics \& Astronomy, Texas A\&M University, College Station, Texas, 77843, U.S.A.}

\author{Alberto Nocera}
\affiliation{Department of Physics and Astronomy and Stewart Blusson Quantum Matter Institute, 
University of British Columbia, Vancouver, B.C., Canada, V6T 1Z1}

\author{Ian Affleck}
\affiliation{Department of Physics and Astronomy and Stewart Blusson Quantum Matter Institute, 
University of British Columbia, Vancouver, B.C., Canada, V6T 1Z1}

\date{\today}

\begin{abstract}

In this work,  we perform a detailed study on the consequences of nonsymmorphic symmetries in the Luttinger phase of the one-dimensional spin-1/2 Kitaev-Heisenberg-Gamma model with an antiferromagnetic Kitaev interaction. 
Nonsymmorphic bosonization formulas for the spin operators are proposed, containing ten non-universal coefficients which are  determined by our  density matrix renormalization group simulations to a high degree of accuracy. 
Using the nonsymmorphic bosonization formulas, 
different Fourier components and decay powers in the correlation functions are disentangled, 
the response to weak magnetic fields is analyzed, and the zigzag magnetic order in two dimensions is recovered from a system of weakly coupled chains. 
We also find a line of critical points with an emergent SU(2)$_1$ conformal symmetry located on the boundary of the Luttinger liquid phase, where a nonabelian version of nonsymmorphic bosonization should be applied. 

\end{abstract}


\maketitle

\section*{INTRODUCTION}

Nonsymmorphic symmetries are symmetry operations involving a combination of rotations or reflections with translations,
where the individual operations do not leave the system invariant on their own \cite{Hiller1986}. 
In recent years, there has been a surge of research interests in studying the roles played by nonsymmorphic symmetries in condensed matter physics.
Although the noninteracting and weakly interacting nonsymmorphic systems  \cite{Liu2014,Fang2015,Wang2016,Alexandradinata2016,Shiozaki2016,Bradlyn2016,Chang2017,Ma2017,Bradlyn2017,Po2017,Wieder2018,Watanabe2018}
have been well-explored
including Dirac insulators, hourglass fermions, and topological semi-metals,
the nonsymmorphic physics in systems with strongly correlations  remain much less investigated  \cite{Lu2017,Thorngren2018}. 

Kitaev materials on the two-dimensional (2D) honeycomb lattice
are strongly correlated systems which
 have  been under intense research attentions in the past decade \cite{Jackeli2009,Chaloupka2010,Singh2010,Price2012,Singh2012,Plumb2014,Kim2015,Winter2016,Baek2017,Leahy2017,Sears2017,Wolter2017,Zheng2017,Rousochatzakis2017,Kasahara2018,Rau2014,Ran2017,Wang2017,Catuneanu2018,Gohlke2018,Liu2011,Chaloupka2013,Johnson2015,Motome2020}, since they can potentially provide solid state realizations for the 2D Kitaev spin-1/2 model known as a platform for topological quantum computations \cite{Kitaev2006,Nayak2008}.
Recently, there have been increasing interests in one-dimensional (1D) versions of Kitaev models \cite{Sela2014,Agrapidis2018,Agrapidis2019,Catuneanu2019,Jiang2019,Yang2020,Yang2020a,Yang2020b,Yang2021b,Yang2022,Luo2021,Luo2021b,Peng2021,You2020,Sorensen2021,Yang2022c,Yang2022e}, with the expectations that such 1D studies can provide hints for the 2D Kitaev physics \cite{Yang2022e}. 

Various 1D Kitaev models have been investigated including Kitaev spin chains, Kitaev spin ladders and doped Kitaev models,
and it has been established that these 1D Kitaev models have intricate nonsymmorphic symmetry group structures \cite{Yang2020,Yang2020a,Yang2022}. 
It is common for such 1D Kitaev models to exhibit gapless Luttinger liquid behaviors, which are useful for understanding the 2D physics,
since the dominant channels of instabilities in the Luttinger liquid theory determine the magnetic orders in 2D \cite{Giamarchi2004}. 
On the other hand, a systematic analysis for the effects of nonsymmorphic symmetries in these Luttinger liquid phases is still lacking. 
 
In this work, we propose the nonsymmorphic bosonization formulas for the spin operators in the framework of Luttinger liquid theory, 
and discuss their applications in analyzing physical properties. 
The Luttinger liquid phase in the spin-1/2 Kitaev-Heisenberg-Gamma ($KJ\Gamma$)  chain with an antiferromagnetic (AFM) Kitaev interaction \cite{Yang2020a} is taken as the example, but the systematic method   is straightforwardly generalizable to other 1D  models with nonsymmorphic symmetries.  
We find that the nonsymmorphic bosonization formulas  contain ten non-universal coefficients,
which are determined to a high degree of accuracy using  density matrix renormalization group (DMRG) simulations.
Once the coefficients are obtained, the nonsymmorphic bosonization formulas can be used to calculate any low energy property of the system.


Three applications of the proposed nonsymmorphic bosonization formulas are discussed. 
As the first application, the critical exponents of the system are extracted with very high precisions.
We note that the fitting curves in previous works for 1D Kitaev models typically involve rugged oscillations \cite{Peng2021}, which hinder a high-precision determination of the critical exponents.
These rugged oscillations indicate a mixture of different Fourier components as well as decay powers in the correlation functions, which can only be disentangled by applying the  nonsymmorphic bosonization formulas. 

As a second application, the response to weak magnetic fields is  analyzed.
Interestingly, the nonsymmorphic bosonization formulas predict  that  the system responds only along a particular fixed direction regardless of the direction of the applied field.
Potential applications of the magnetic field response in spintronics and quantum information processing are discussed. 
As the third application, a system of weakly coupled spin-1/2 $KJ\Gamma$ chains on the honeycomb lattice with an AFM Kitaev interaction is studied by treating the interchain interactions in a self-consistent mean field approach. 
The magnetic pattern produced by the nonsymmorphic bosonization formulas recovers the zigzag order in 2D, previously obtained by classical analysis  and the exact diagonalization method \cite{Rau2014}.

Finally,  we also reveal  a line of  critical points with an emergent SU(2)$_1$ conformal symmetry located on the boundary of the Luttinger liquid phase.
The nonsymmorphic bosonization formulas equally apply to this line, except that nonabelian bosonization rather than abelian bosonization should be used.

\section*{RESULTS}

\subsection*{Model and symmetries\\}

\begin{figure}[h]
\includegraphics[width=8cm]{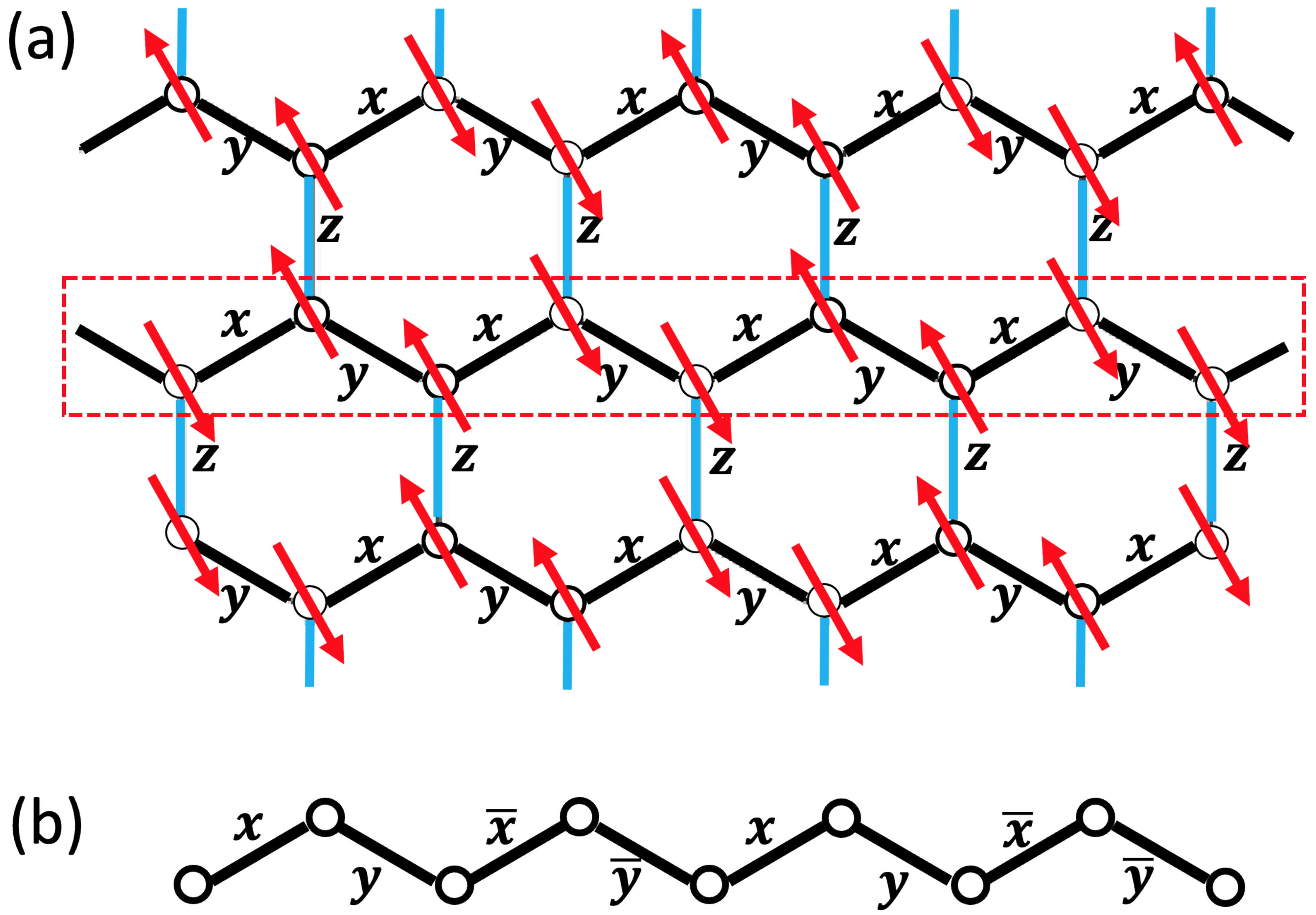}
\caption{(a) Zigzag order of the spin-1/2 anisotropic  $KJ\Gamma$ model on the honeycomb lattice;
(b) bond pattern after the four-sublattice rotation for the chain   in (a)  enclosed by the red dashed line.
In (a), the black and blue lines denote the strong and weak bonds;
the red arrows represent the directions of the spin orientations which  are along the $\pm\frac{1}{\sqrt{(a_C)^2+(b_C)^2+(c_C)^2}}(a_C,-b_C,c_C)$ direction, where $a_C$, $b_C$, and $c_C$ are  coefficients in the nonsymmorphic bosonization formulas. 
} \label{fig:bond_pattern}
\end{figure}

The 1D spin-1/2 $KJ\Gamma$  model is defined as
\begin{eqnarray}
H=\sum_{\langle ij\rangle=\gamma}\big[ KS_i^\gamma S_j^\gamma+ J\vec{S}_i\cdot \vec{S}_j+\Gamma (S_i^\alpha S_j^\beta+S_i^\beta S_j^\alpha)\big],
\label{eq:Ham}
\end{eqnarray}
in which $\gamma$ is the spin direction associated with the bond between nearest neighboring sites $i$ and $j$, alternating between $x$ and $y$ as shown in Fig. \ref{fig:bond_pattern} (a); 
$\alpha\neq \beta$ are the two spin directions among $x,y,z$  other than $\gamma$; 
and $K,J,\Gamma$ are the Kitaev, Heisenberg and Gamma couplings, respectively.
The Hamiltonian in Eq. (\ref{eq:Ham}) is obtained by selecting one row out of the 2D $KJ\Gamma$ model on the honeycomb lattice as shown in Fig. \ref{fig:bond_pattern} (a).
We will consider the parameter region $K>0$, $J<0$.
Since $\Gamma$ changes sign under a global spin rotation around $z$-axis by $\pi$, there is the equivalence $(K,J,\Gamma)\simeq (K,J,-\Gamma)$, and we will take $\Gamma>0$ for simplicity. 

A useful transformation is the four-sublattice rotation $U_4$ \cite{Chaloupka2013,Yang2020a},
which leaves the spins on sites  $4n$ ($n\in\mathbb{Z}$) unchanged, and acts as 
$\pi$-rotations around $y$-, $z$-, and $x$-axes on sites $1+4n$, $2+4n$ and $3+4n$, respectively.
The transformed Hamiltonian acquires the following form
\begin{eqnarray}
H^{\prime}=\sum_{\langle ij\rangle}\big[ K^\prime S_i^\gamma S_j^\gamma-J \vec{S}_i\cdot \vec{S}_j +\epsilon(\gamma) \Gamma (S_i^\alpha S_j^\beta+S_i^\beta S_j^\alpha)\big],
\label{eq:4rotated}
\end{eqnarray}
in which $K^\prime=K+2J$; the bond pattern for $\gamma\in\{x,y,\bar{x},\bar{y}\}$ is shown in Fig. \ref{fig:bond_pattern} (b);
$S_j^{\bar{\gamma}}=S_j^\gamma$; 
$\alpha\neq \beta$ are the two spin directions different from $\gamma$; 
and $\epsilon(x)=\epsilon(y)=-\epsilon(\bar{x})=-\epsilon(\bar{y})=1$.
The advantage of $U_4$ is that it reveals a hidden SU(2) symmetric AFM point at $J=-K/2$, $\Gamma=0$, which provides a perturbative starting point to analyze the parameter region nearby \cite{Chaloupka2013,Yang2020a}. 
The system before and after the $U_4$ transformation will be termed as original and four-sublattice rotated frames, respectively. 

\begin{figure}[h]
\includegraphics[width=4cm]{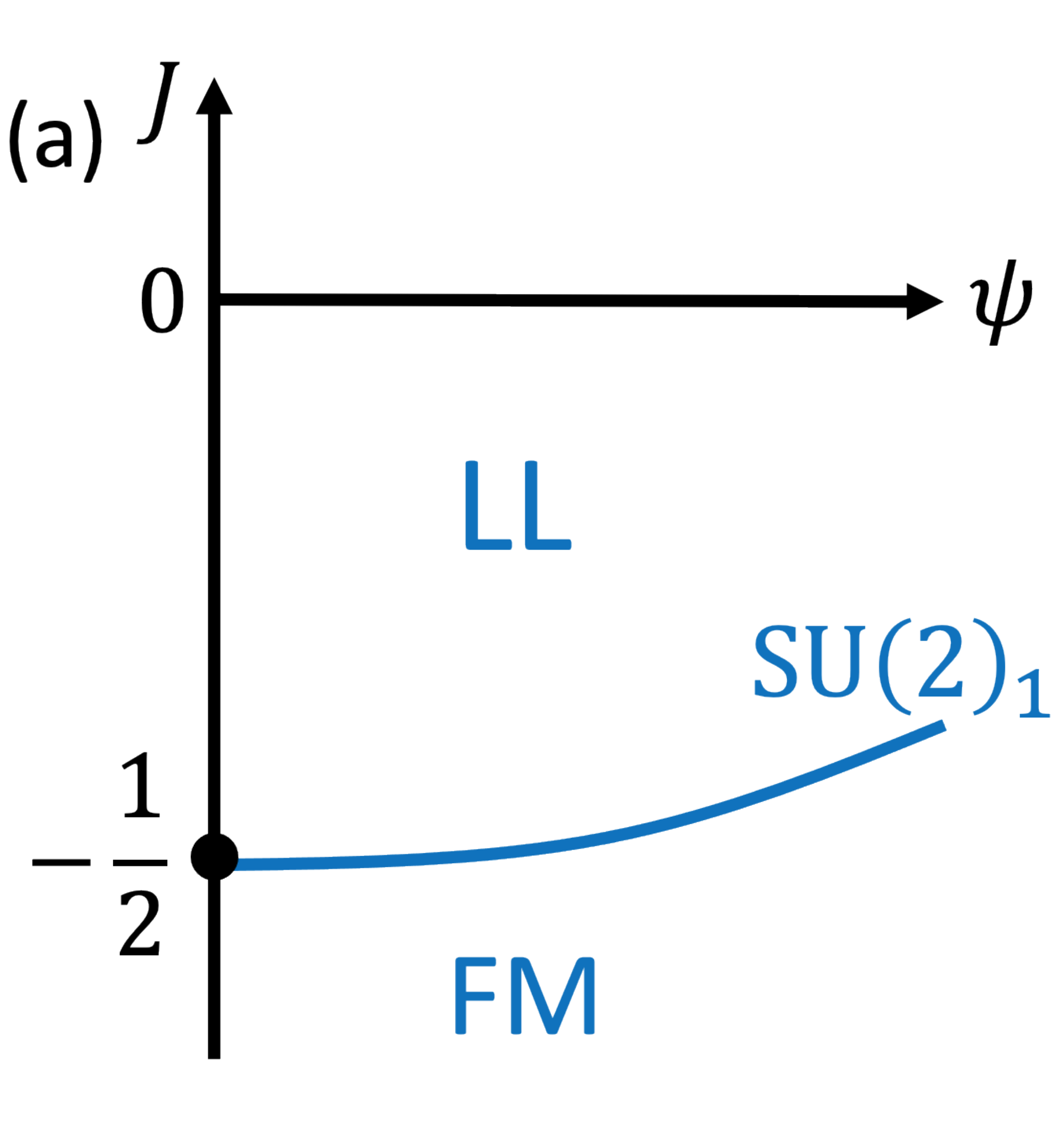}
\includegraphics[width=4cm]{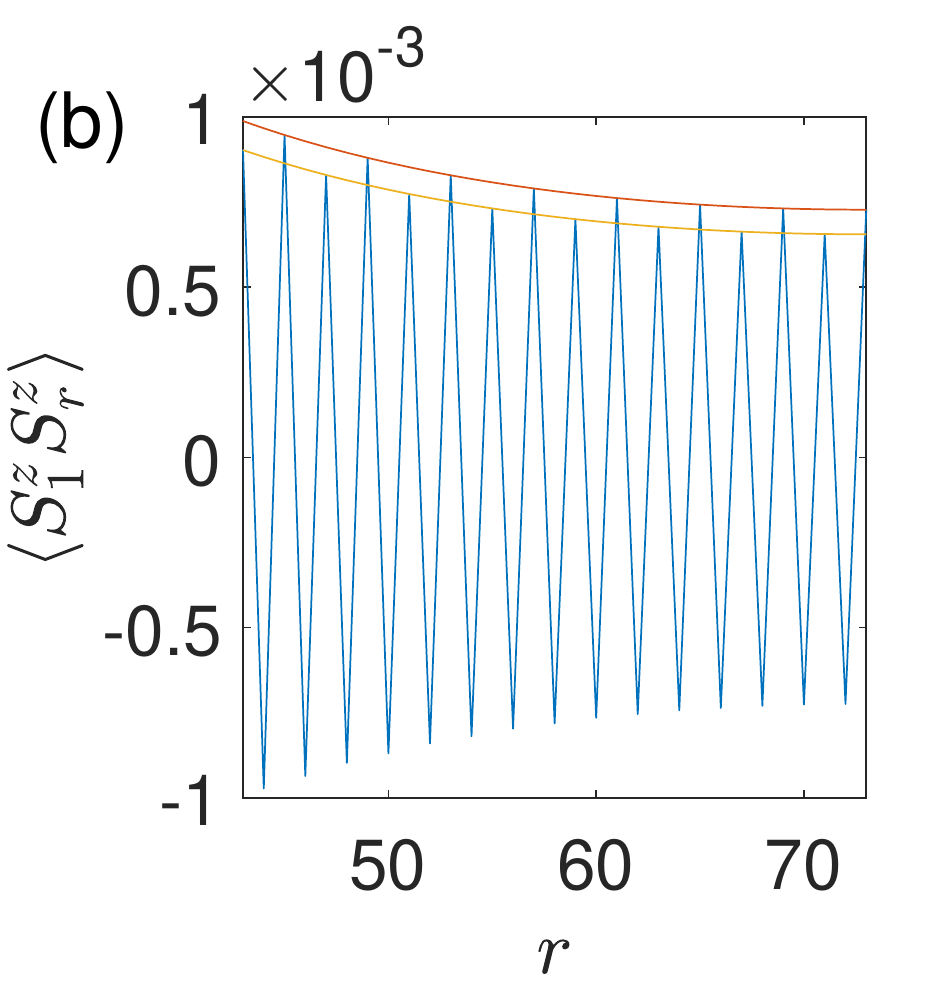}
\caption{(a) Phase diagram of the spin-1/2 $KJ\Gamma$ chain in the region $K>0,J<0$, in which the vertical axis is $J$ and the horizontal axis is $\psi$ where $K=\cos(\psi)$ and $\Gamma=\sin(\psi)$;
(b) $\langle S_1^zS_{1+r}^z \rangle$ as a function of $r$.
In (a), ``LL", ``FM", and ``SU(2)$_1$" denote the Luttinger liquid phase, the FM phase, and the line of points with emergent SU(2)$_1$ conformal symmetry, respectively.
In (b), DMRG numerics are  performed for $K^\prime=1,J=-1,\Gamma=0.35$ in Eq. (\ref{eq:4rotated}) on a system of $L=144$ sites with periodic boundary conditions.
} \label{fig:phase}
\end{figure}

The phase diagram in the $K>0,J<0$ region is  shown in Fig. \ref{fig:phase} (a) \cite{Yang2020a}, in which the horizontal axis $\psi$ is defined by $K=\cos(\psi),\Gamma=\sin(\psi)$.
We will focus on the Luttinger liquid phase (denoted as ``LL" in Fig. \ref{fig:phase} (a)) in this work.
When $K^\prime>0$ and $\Gamma\neq 0$, at low energies, the $K^\prime$ term in $H^\prime$ contributes to an easy-plane anisotropy whereas the $\Gamma$ term cancels to leading order since the differences between neighboring sites are smeared out in the long distance limit,
which provides an intuitive explanation for the origin of the Luttinger liquid phase.
More rigorously, the existence of the Luttinger liquid phase can be established  by analyzing the symmetry allowed terms in a field theory perturbation of the hidden SU(2) symmetric point $(\psi=0,J=-1/2)$ \cite{Yang2020a}. 

For later use, we briefly discuss the symmetries of $H^\prime$ in Eq. (\ref{eq:4rotated}) \cite{Yang2020a}.
The symmetry group $G$ of $H^\prime$ is generated by time reversal symmetry $T$, screw operation $R(\hat{z},-\pi/2)T_a$, and composite operation $R(\hat{y},\pi)I$, where $a$ is the lattice constant, $T_{na}$ is the translation operator by $n$ lattice sites, $I$ is the spatial inversion with respect to the bond center between sites $2$ and $3$,
and $R(\hat{n},\phi)$ represents a global spin rotation around $\hat{n}$-direction by an angle $\phi$.
The group structure of $G$ satisfies $G/\mathopen{<}T_{4a}\mathclose{>}\cong D_{4h}$ in which $\mathopen{<}...\mathclose{>}$ is the group generated by the elements within the brackets, and $D_{4h}\cong D_4\times \mathbb{Z}_2^T$ where $D_4$ is the dihedral group of order eight and $\mathbb{Z}_2^T$ is the $\mathbb{Z}_2$ group generated by $T$.
Notice that although $[R(\hat{z},-\pi/2)T_a]^4$ is the identity element in the quotient group $G/\mathopen{<}T_{4a}\mathclose{>}$, it is not the identity in $G$, which is the reason why $G$ is a nonsymmorphic group.
In a mathematically rigorous language, $G$ is nonsymmorphic since the short exact sequence $1\rightarrow \mathopen{<}T_{4a}\mathclose{>}\rightarrow G\rightarrow D_{4h}\rightarrow 1$ is non-splitting. 
More detailed discussions of the nonsymmorphic structure of $G$ are included in Sec. II in Supplementary Materials (SM)  \cite{SM}.


\subsection*{Nonsymmorphic bosonization formulas\\}

\begin{figure*}[ht]
\includegraphics[width=13cm]{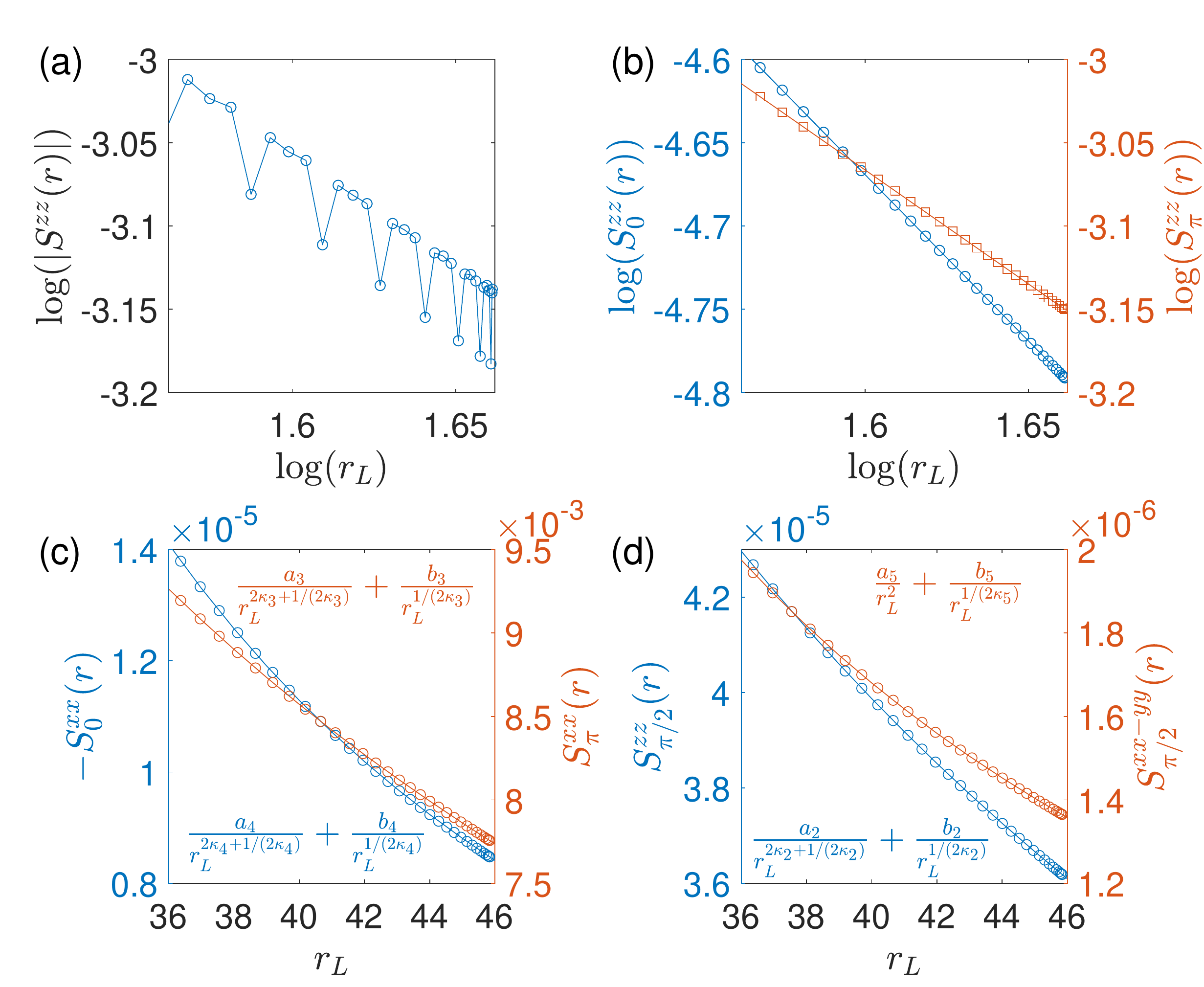}
\caption{
(a) $|S^{zz}(r)|$ as a function of $r_L=L/\pi\sin(\pi r/L)$ on a log-log scale where $S^{zz}(r)=\langle S_1^z S_{1+r}^z\rangle$;
(b) fits of $-S^{zz}_0(r)$ (blue) and $S^{zz}_\pi(r)$ (orange) as functions of $r_L$ on a log-log scale using linear relations; 
(c) fits of $-S^{xx}_0(r)$ (blue) and $S^{xx}_\pi(r)$ (orange);  
(d) fits of $S^{zz}_{\pi/2}$ (blue) and $S^{xx-yy}_{\pi/2}$ (orange).
In  (c,d), the functional forms used for the fits are shown.
DMRG numerics are  performed for $K^\prime=1,J=-1,\Gamma=0.35$ in the four-sublattice rotated frame.
} \label{fig:correlations}
\end{figure*}

Throughout this section, the analysis of nonsymmorphic bosonization formulas will be carried out within the four-sublattice rotated frame.
In the Luttinger liquid phase, the low energy physics is described by the Luttinger liquid Hamiltonian \cite{Giamarchi2004,Affleck1988}
\begin{eqnarray}
H_{LL}=\frac{v}{2} \int dx [\kappa^{-1} (\nabla \varphi)^2 +\kappa (\nabla \theta)^2],
\label{eq:H_LL}
\end{eqnarray}
in which $v$ is the velocity, $\kappa$ is the Luttinger parameter,
and the $\theta,\varphi$ fields satisfy the commutation relation $[\varphi(x),\theta(x^\prime)]=\frac{i}{2}\text{sgn}(x^\prime-x)$.
We propose the bosonization formulas which are consistent with the nonsymmorphic symmetry group of the system.
In general, the spin operators in the four-sublattice rotated frame can be related to the low energy fields as
\begin{eqnarray}
S^{\alpha}_{j+4n} = D^{(j)}_{\alpha\beta} J^{\beta}(x)+(-)^j C^{(j)}_{\alpha\beta} N^{\beta}(x),
\label{eq:bosonization_fourier}
\end{eqnarray}
in which $\alpha,\beta=x,y,z$; $1\leq j\leq 4$; $x=(j+4n)a$;  $D^{(j)},C^{(j)}$ are $3\times 3$ matrices;
the smeared spin densities $J^\alpha,N^\alpha$ ($\alpha=x,y,z$) are given by
$J^\pm=\frac{2}{a}\cos(\sqrt{4\pi}\varphi)e^{\pm i\sqrt{\pi}\theta}$,  $J^z=-\sqrt{2}\pi \nabla \varphi$, 
$N^\pm=\frac{\sqrt{2}}{a} e^{\pm i \sqrt{\pi}\theta}$,   $N^z=\frac{\sqrt{2}}{a}\sin(\sqrt{4\pi}\varphi)$, 
where $J^\pm=J^x\pm iJ^y$ and $N^\pm=N^x\pm iN^y$.
The invariance under the symmetries $R(\hat{z},-\pi/2)T_a$ and $R(\hat{y},\pi)I$ requires
\begin{eqnarray}
&D^{(j+1)}=R_z D^{(j)} (R_z)^{-1},~  D^{(5-j)}=R_y D^{(j)}(R_y)^{-1},\nonumber\\
&C^{(j+1)}=R_z C^{(j)} (R_z)^{-1},~  C^{(5-j)}=R_y C^{(j)}(R_y)^{-1},
\label{eq:constraints}
\end{eqnarray}
in which $j=5$ is understood as $1$ (modulo $4$);  $R_z$ and $R_y$ are the $3\times 3$ rotation matrices corresponding to $R(\hat{z},-\pi/2)$ and $R(\hat{y},\pi)$, respectively. 
The solution of Eq. (\ref{eq:constraints})  is 
\begin{flalign}
C^{(1)}=\left(\begin{array}{ccc}
a_C & b_C & c_C\\
b_C & a_C & -c_C\\
h_C & -h_C & i_C
\end{array}\right), 
D^{(1)}=\left(\begin{array}{ccc}
a_D & b_D & c_D\\
b_D & a_D & -c_D\\
h_D & -h_D & i_D
\end{array}\right), 
\label{eq:coeff_matrix}
\end{flalign}
and the remaining matrices $C^{(j)},D^{(j)}$ ($j=2,3,4$) can be obtained from the relations in Eq. (\ref{eq:constraints}). 

Here we comment  on the differences from the conventional U(1) symmetric cases.
The low energy field theory in Eq. (\ref{eq:H_LL}) has an emergent U(1) symmetry $\theta\rightarrow \theta+\beta$, corresponding to rotations around $z$-axis. 
When the microscopic Hamiltonian is U(1) invariant, the bosonization formulas of the spin operators are given by 
$S_j^\alpha=\lambda J^\alpha+\mu(-)^j  N^\alpha$, in which $\lambda,\mu$ are constants. 
However, these relations cease to apply in the $KJ\Gamma$ chain.
In the present nonsymmorphic case,
the bosonization  formulas in Eq. (\ref{eq:bosonization_fourier}) have low energy spinon modes at $\pm\pi/2$ wavevectors in addition to zero and $\pi$-wavevectors, and
$S_j^\lambda$ contains cross-directional components $J^\mu,N^\mu$ where $\mu\neq \lambda$, not just diagonal ones.

Next we discuss the numerical results for the spin correlation functions.
DMRG simulations are performed in the four-sublattice rotated frame at a representative point in the Luttinger liquid phase with the parameters $K^\prime=1,J=-1,\Gamma=0.35$, on a system of $L=144$ sites using periodic boundary conditions.
Fig. \ref{fig:phase} (b) displays $\langle S_1^zS_{1+r}^z \rangle$ as a function of $r$, which is dominated by the $\pi$-wavevector oscillation;
however, as can be seen from the data points guided by the red and orange lines, there exists a delicate subdominant structure having a four-site periodicity.  
Fig. \ref{fig:correlations} (a) shows $|\langle S_1^zS_{1+r}^z \rangle|$ as a function of $r_L$ on a log-log scale, where $r_L=L/\pi\sin(\pi r/L)$ in accordance with conformal field theory in finite size systems \cite{Affleck1988}. 
It can be clearly seen that Fig. \ref{fig:correlations} (a) contains rugged oscillations in the data points, 
indicating an entangling of different Fourier components,
which hinders a high-precision  determination of the critical exponent. 

To study the spin correlation functions in more details, 
we apply the nonsymmorphic bosonization formulas such that  different Fourier components and decay powers can be disentangled.
Notice that  Eq. (\ref{eq:bosonization_fourier}) predicts
$\langle S_1^zS_{1+r}^z \rangle=S^{zz}_0(r)+(-)^{r-1}S^{zz}_\pi(r)-2\cos(\frac{\pi}{2}(r+\frac{1}{2}))S^{zz}_{\pi/2}(r)$,
where $S^{zz}_0(r)=-i_D^2r^{-2}$, $S^{zz}_{\pi}(r)=i_C^2r^{-2\kappa}$, and $S^{zz}_{\pi/2}(r)=c_C^2r^{-1/(2\kappa)}-c_D^2r^{-[2\kappa+1/(2\kappa)]}$.
The blue and orange lines in Fig. \ref{fig:correlations} (b) show the fits of $-S^{zz}_0$ and $S^{zz}_\pi$ as a function of $r_L$  on a log-log scale, respectively, using linear relations, 
where $r$ is replaced by $r_L=L/\pi\sin(\pi r/L)$.
The slope of the blue line is $-1.995$, very close to the predicted value of $-2$;
the slope of the orange line yields a Luttinger parameter $\kappa_1=0.683$.
The fit of $S^{zz}_{\pi/2}(r)$ using a functional form $a_2r_L^{-1/(2\kappa_2)}+b_2r_L^{-[2\kappa_2+1/(2\kappa_2)]}$ is shown as the blue line in Fig. \ref{fig:correlations} (d) which gives $\kappa_2=0.685$.
The extracted values of $|i_\Lambda|$ and $|c_\Lambda|$ ($\Lambda=C,D$) from Fig. \ref{fig:correlations} (b,d) are included in Table \ref{table:coeff}.

For the  transverse correlations, Eq. (\ref{eq:bosonization_fourier}) predicts 
the zero- and $\pi$-wavector oscillating components of $\langle S_1^xS_{1+r}^x \rangle$ to be
$S^{xx}_0(r)=-a_D^2 r^{-2\kappa-1/(2\kappa)}+b_C^2/r^{-1/(2\kappa)}$ and
$S^{xx}_\pi(r)=-b_D^2r^{-2\kappa-1/(2\kappa)}+a_C^2r^{1/(2\kappa)}$, respectively.
The blue and orange lines in Fig. \ref{fig:correlations} (c) show the fits of $-S^{xx}_0$ and $S^{xx}_\pi$,  
which give $\kappa_3=0.681$ from $S^{xx}_0$, and $\kappa_4=0.681$ from $S^{xx}_\pi$.  
In addition, Eq. (\ref{eq:bosonization_fourier}) predicts the $\pm \pi/2$-component of $\langle S_1^xS_{1+r}^x \rangle-\langle S_1^yS_{1+r}^y \rangle$ to be $-\cos(\frac{\pi}{2}(r+\frac{1}{2}))S^{xx-yy}_{\pi/2}$ where $S^{xx-yy}_{\pi/2}=h_D^2 r^{-2}+h_C^2r^{-2\kappa}$.
The fit for $S^{xx-yy}_{\pi/2}$ is shown as the orange line in Fig. \ref{fig:correlations} (d),   giving $\kappa_5=0.682$.
The extracted values of $|a_\Lambda|, |b_\Lambda|,|h_\Lambda|$ ($\Lambda=C,D$) from Fig. \ref{fig:correlations} (c,d) are included in Table \ref{table:coeff},
where the only exception is $|b_D|$.
We note that the $b_D$ term decays much faster than the $a_C$ term, since the exponents are $2\kappa+1/(2\kappa)$ for the former and $1/(2\kappa)$ for the latter. In addition, $|b_D|$ itself is much smaller than $|a_C|$. These two effects make it very difficult to extract a reliable value of $|b_D|$ since the $b_D$ and $a_C$ terms are mixed in $S^{xx}_\pi(r)$.

\begin{table}[!t]
\centering
 \begin{tabular}{| c || c | c | c | c | c | } 
 \hline
   & $|a_\Lambda|$ & $|i_\Lambda|$ & $|c_\Lambda|$ & $|h_\Lambda|$ & $|b_\Lambda|$ \\
 \hline
$\Lambda=C$ &  0.129 & 0.363 & 0.0244 & 0.0138 & 0.00103 \\
\hline
$\Lambda=D$ & 0.161 & 0.182 & 0.0359 & 0.0266 & ? \\
 \hline
\end{tabular}
\caption{Extracted  values of $|w_\Lambda|$ ($w=a,i,c,h,b$; $\Lambda=C,D$)
for the representative point $K^\prime=1,J=-1,\Gamma=0.35$,
in which $|b_D|$ is too small and a reliable value cannot be extracted.
}
\label{table:coeff}
\end{table}

From on the above discussions, we see that with the help of the proposed nonsymmorphic bosonization formulas, the fitting curves are rendered smooth without any ruggedness,
and nearly all the bosonization coefficients are numerically determined which can be used to calculate any low energy property of the system. 
In particular, different Fourier components of the correlation functions give five independently extracted values  of Luttinger parameters, denoted as $\kappa_i$ ($1\leq i\leq 5$) above, which match with each other within an accuracy of $0.6\%$,
indicating a consistency  of the theory and a high precision of the numerics. 
We note that the systematic method of nonsymmorphic bosonization discussed in this section can be directly generalized to other 1D nonsymmorphic systems including Kitaev ladders and doped Kitaev models,
which is helpful for understanding more delicate structures previously ignored in those cases. 



\subsection*{Response to magnetic field\\}

\begin{figure*}[ht]
\begin{center}
\includegraphics[width=5.8cm]{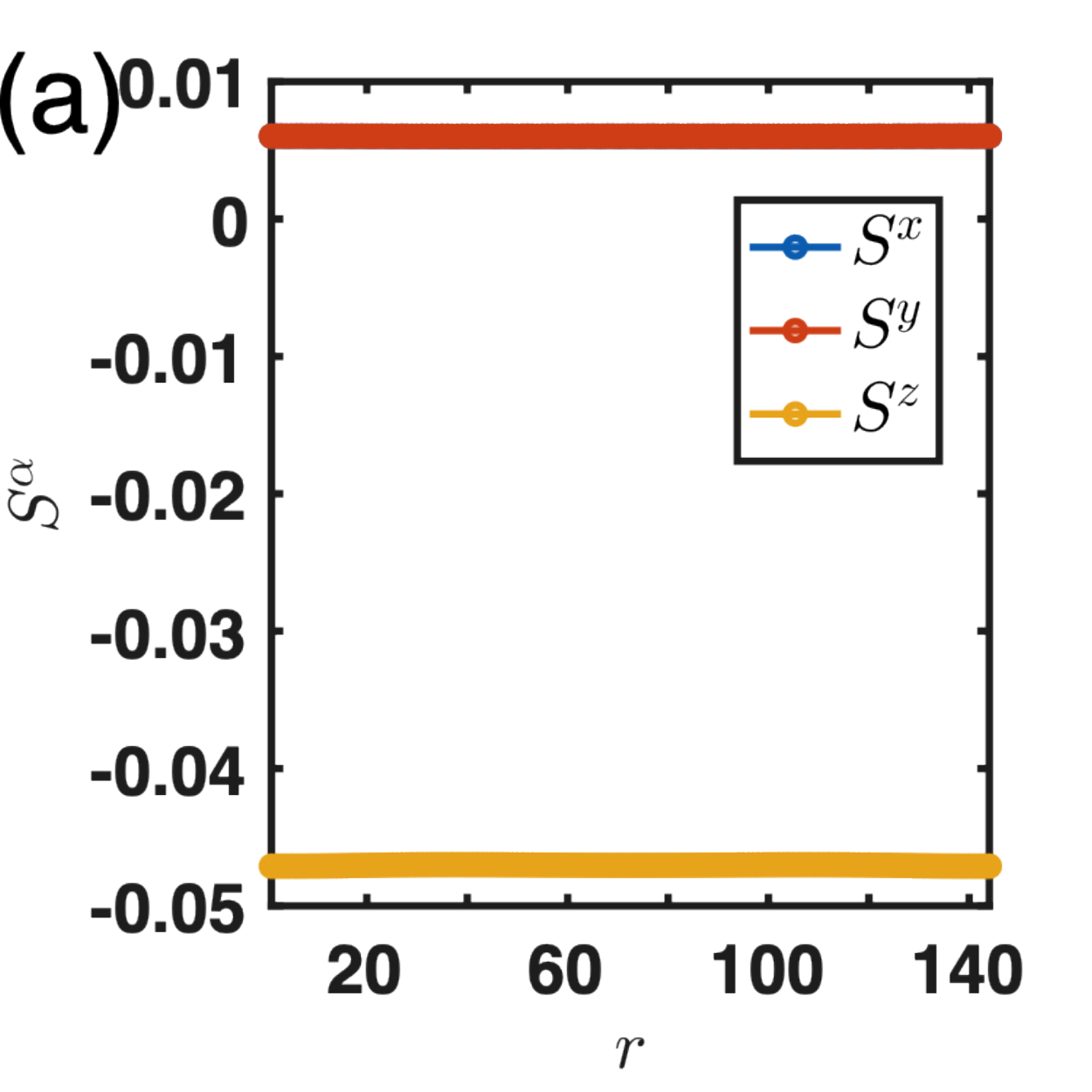}
\includegraphics[width=6cm]{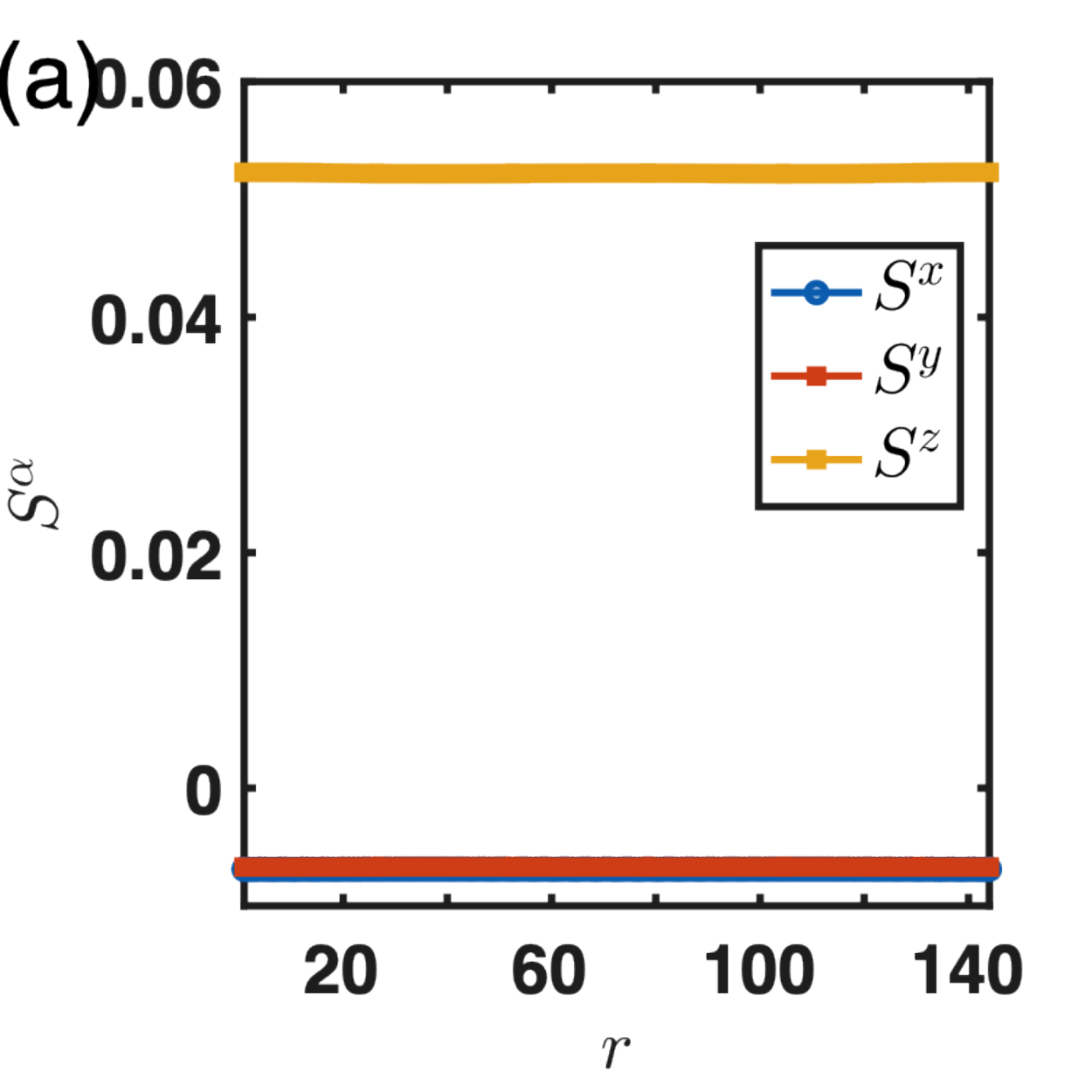}
\caption{Spin expectation values of $S_r^\alpha$ ($\alpha=x,y,z$) under uniform magnetic fields (a) $h^z=10^{-4}$ along $z$-direction, (b) $h^x=10^{-3}$, as functions of site $r$ ($1 \leq r \leq 144$).
The data of $S^x$ and $S^y$ nearly overlap, making the points of $S^x$ hardly identifiable. 
DMRG simulations are performed for $K=2.1$, $J=-1$, $\Gamma=0.8$, on a system of $L=144$ sites, using periodic boundary conditions. 
} \label{fig:h_response}
\end{center}
\end{figure*}

As an application of Eq. (\ref{eq:bosonization_fourier}), we consider the response of the 1D $KJ\Gamma$ model to weak magnetic fields.
By applying $(U_4)^{-1}$ to Eq. (\ref{eq:bosonization_fourier}), it can be shown that at low energies, the coupling to a uniform weak magnetic field in the original frame is given by 
\begin{eqnarray}
\sum_j \vec{h}\cdot \vec{S}_j=4[h_C(h^x+h^y)+i_Ch^z]\int dx N^z.
\end{eqnarray}
Since $N^z$ is a relevant perturbation, a spin gap opens for an arbitrarily small uniform field,
and  the spin magnetization can be determined from Eq. (\ref{eq:bosonization_fourier}) as $\langle \vec{S}_j\rangle=\langle N^z\rangle(h_C,h_C,i_C)$ in the original  frame.
Interestingly, in the weak field limit, the direction of the spin magnetization is fixed by $h_c$ and $i_c$ up to an overall sign and  does not depend on the direction of the applied magnetic field. 

To verify this prediction, we study the response to magnetic fields using DMRG at a representative point  $K=2.1$, $J=-1$, $\Gamma=0.8$ in the original frame.
Fig. \ref{fig:h_response} (a) and (b) show $\langle \vec{S}_r \rangle$ as a function of $r$ under uniform magnetic fields $h^z=10^{-4}$ along $z$-direction and $h^x=10^{-3}$ along $x$-direction, respectively. 
The extracted value $h_C/i_C = \langle S^x_r\rangle /\langle S^z_r \rangle $ is $-0.128$ (for $h^z$) and $-0.130$ (for $h^x$), which are consistent with each other.
We note that the response behavior of a fixed magnetization direction regardless of the direction of the applied field is a special property which can only be captured by the nonsymmorphic bosonization formulas and does not appear in U(1) symmetric systems. 

For temperatures smaller than the spin gap, the above discussion shows that a ring-shape AFM $KJ\Gamma$ system can display slow relaxation in magnetization dynamics under external  magnetic fields. 
If realized in real materials, this behavior could find natural applications in single molecule magnetic cluster systems, which hold promise as tunable components for spintronic devices as well as quantum information storage and quantum computation \cite{Meier2003,GaitaArino2019}.


\subsection*{2D zigzag order\\}

As another application of Eq. (\ref{eq:bosonization_fourier}), 
we derive the pattern of spin ordering for a system of weakly coupled 1D chains as shown in Fig. \ref{fig:bond_pattern} (a), 
which is an anisotropic spin-1/2  $KJ\Gamma$ model on the  honeycomb lattice.
The interaction on bond $\gamma$ is 
\begin{eqnarray}
\alpha_\gamma[KS_i^\gamma S_j^\gamma+ J\vec{S}_i\cdot \vec{S}_j+\Gamma (S_i^\alpha S_j^\beta+S_i^\beta S_j^\alpha)],
\end{eqnarray}
where $\alpha_x=\alpha_y=1$ and $\alpha_z\ll 1$.
Using the nonsymmorphic bosonization formulas, the mean field Hamiltonian in the four-sublattice rotated frame for the chain enclosed by the red dashed line in Fig. \ref{fig:bond_pattern} (a) can be derived as 
\begin{eqnarray}
H_{MF}=H_{LL}-\frac{\lambda}{a^3} \langle \cos(\sqrt{\pi}\theta) \rangle \int dx \cos(\sqrt{\pi}\theta),
\end{eqnarray}
where $H_{LL}$ is the Luttinger liquid Hamiltonian in Eq. (\ref{eq:H_LL}) and $\lambda=  |J|(a_C^2+b_C^2)-(K+J) c_C^2+2\Gamma a_Cb_C$.
Since $N^x$ and $N^y$ have the smallest scaling dimension (both equal to $(4\kappa)^{-1}$), it is expected that the leading instability is in the $S^xS^y$-plane.
Although different directions in the $S^xS^y$-plane in the Luttinger liquid theory are equivalent, the symmetry allowed  irrelevant term $\cos(4\sqrt{\pi}\theta)$  stabilizes a spin order which satisfies the condition that one of $\langle N^\alpha \rangle$ ($\alpha=x,y$) does not vanish, but not both. 

Assuming $\langle N^x \rangle\neq 0$, 
applying  Eq. (\ref{eq:bosonization_fourier}), and transforming back to the original frame, 
the spin expectation values are determined as
$\vec{S}_{m,n}=\sqrt{2}\cos(\frac{\pi}{2}(n-m)+\frac{1}{2})\langle N^x \rangle (a_C,-b_C,c_C)$, in which $m$ and $n$ are the row and column indices in the equivalent brick wall lattice, respectively, and $m=1$ for the enclosed chain in Fig. \ref{fig:bond_pattern} (a).
The red arrows in Fig. \ref{fig:bond_pattern} (a) represent directions of the spin expectation values,
which are along the $\pm\frac{1}{\sqrt{(a_C)^2+(b_C)^2+(c_C)^2}}(a_C,-b_C,c_C)$ direction. 
It can be seen that the pattern in  Fig. \ref{fig:bond_pattern} (a) is exactly the 2D zigzag order, 
recovering the zigzag phase in the 2D $KJ\Gamma$ model \cite{Chaloupka2013,Rau2014}.
It is worth to note that the nonsymmorphic bosonization formulas are crucial to obtain a zigzag order with a variable direction, 
since the spin direction will be otherwise fixed along $x$-direction if the conventional U(1) symmetric bosonization formulas are used. 

Furthermore, using the variational method for the massive 1D sine-Gordon model \cite{Giamarchi2004},
the spin gap $\Delta$ and the expectation value $\langle N^x \rangle$ can be  mutually determined as
$\Delta = v\Lambda \big[\pi\lambda\langle \cos(\sqrt{\pi}\theta) \rangle/(v\kappa \Lambda^2a^3)\big]^{4\kappa/(8\kappa-1)}$
and $\langle \cos(\sqrt{\pi}\theta) \rangle=[\Delta/(v\Lambda)]^{1/(4\kappa)}$, where $\Lambda\sim 1/a$ is the ultra-violet cutoff for the Luttinger liquid theory.
Self-consistency then requires
$\langle \cos(\sqrt{\pi}\theta) \rangle =\big[\pi\lambda/(v\kappa \Lambda^2a^3)\big]^{\frac{1}{8\kappa-2}}$ which  determines the strength of the spin magnetization.
Detailed discussions on weakly coupled chains are included in Sec. VI in SM \cite{SM}.

\subsection*{Line of points with emergent SU(2)$_1$ conformal invariance\\}

\begin{figure}[ht]
\begin{center}
\includegraphics[width=7cm]{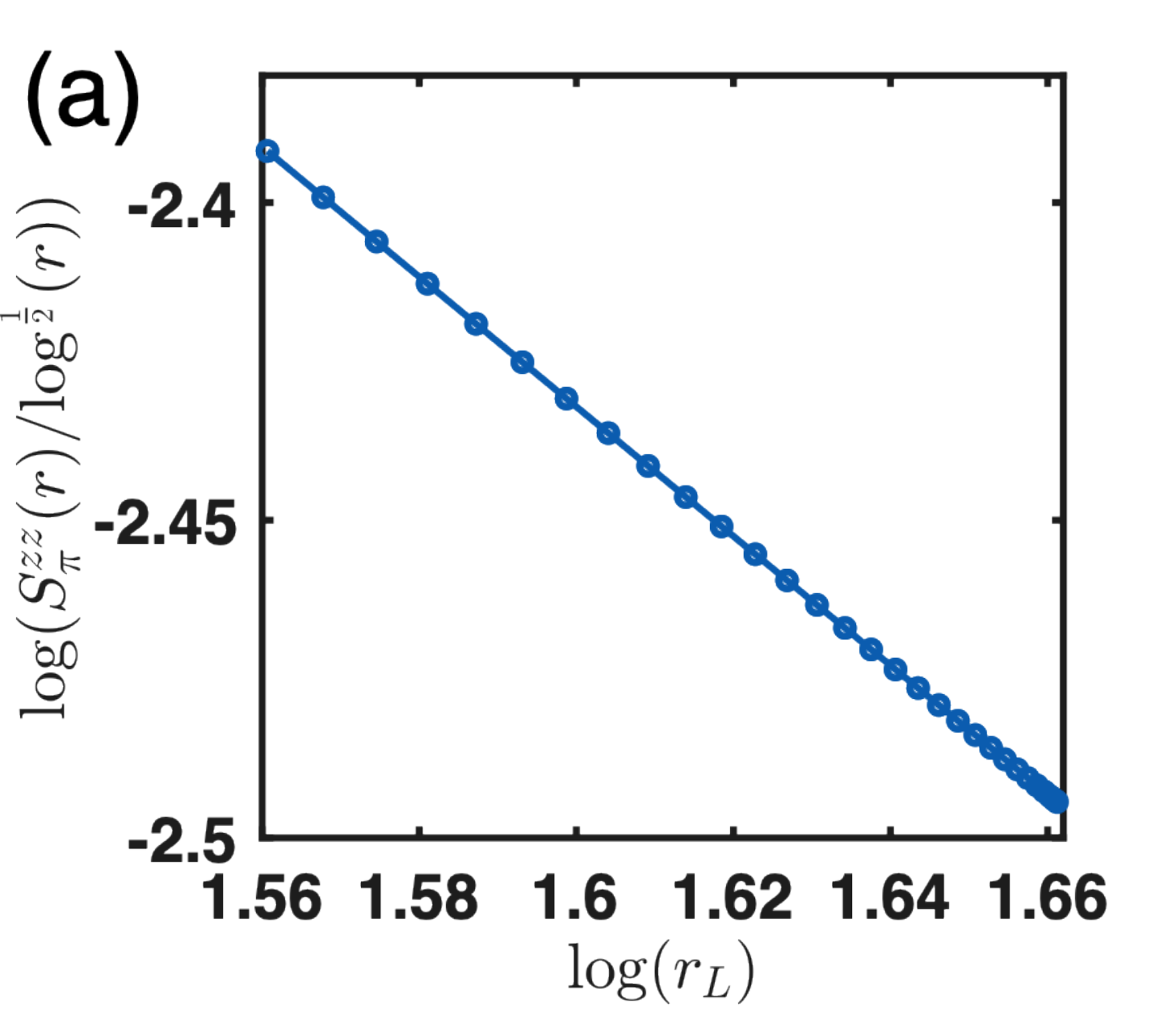}
\caption{
$S^{zz}_\pi (r)/\sqrt{\log_{10}(r_L)}$ as a function of $r_L$ on a log-log scale, where the slope is extracted to be $-1.019$.
DMRG simulations are performed for $K=2$, $J=-1$, $\Gamma=0.35$, on a system of $L=144$ sites, using periodic boundary conditions. 
} \label{fig:SU2_1}
\end{center}
\end{figure}

Finally, although the nonsymmorphic group is discrete and planar, 
we demonstrate that the phase boundary between the LL and FM phases in Fig. \ref{fig:phase} has an emergent SU(2)$_1$ conformal symmetry.
The FM phase  corresponds to a Neel order in the four-sublattice rotated frame with $\langle N^z\rangle\neq0$.
Hence, at low energies, the phase transition between LL and FM phases  is  the same  as the easy-plane to easy-axis transition in the XXZ model, which is in the universality class described by the SU(2)$_1$ Wess-Zumino-Witten (WZW) model.
Indeed, as can be seen from Eq. (\ref{eq:4rotated}), when $K+2J=0$, the system is SU(2) invariant at long distances regardless of the value of $\Gamma$, since the $\Gamma$ term vanishes when the neighboring sites get smeared and can no longer be distinguished.
Higher order effects shift the location of the phase transition line, but the emergent SU(2)$_1$ symmetry is expected to persist. 

Instead of abelian bosonization, the nonabelian bosonization should be applied to the SU(2)$_1$ line, 
i.e., $J^\alpha$ and $N^\alpha$ ($\alpha=x,y,z$) in Eq. (\ref{eq:bosonization_fourier}) should be replaced by the current operators and primary fields in the SU(2)$_1$ WZW theory, respectively. 
Nevertheless, the nonsymmorphic nonabelian bosonization formulas on the SU(2)$_1$ line remain the same form as
Eq. (\ref{eq:bosonization_fourier}), still containing ten free coefficients.
The SU(2)$_1$ WZW theory together with the nonsymmorphic nonabelian bosonization predicts $\pi$ wavevector component of  $\langle S_1^z S_{r}^z \rangle$ to be $S^{zz}_\pi(r)=i_C^2\frac{\ln^{1/2}(r/r_0)}{r}$, 
in which the logarithmic correction comes from the marginal operator $\vec{J}_L\cdot\vec{J}_R$ in the low energy theory,
and $r_0$ is a microscopic cutoff. 
For numerical verifications, DMRG simulations are performed on a system of $L=144$ sites using periodic boundary conditions
at $K=2$, $J=-1$, $\Gamma=0.35$ (such that $K+2J=0$).
Fig. \ref{fig:SU2_1} (a) shows $S^{zz}_\pi (r)/\sqrt{\log_{10}(r_L)}$ as a function of $r_L$ on a log-log scale,
giving a slope $-1.019$, very close to the predicted value of $-1$.

\section*{DISCUSSION}

In summary, we have systematically investigated the roles played by the nonsymmorphic symmetries in the low energy Luttinger liquid theory in 1D generalized Kitaev spin-1/2 model. 
The nonsymmorphic bosonization formulas are proposed, which are crucial for gaining an understanding on many physical properties. 
As an application of  the nonsymmorphic bosonization method, different Fourier components and decay powers in the correlation functions are disentangled, enabling a removal of the ruggedness in the fitting curves so that high-precision results of the critical exponents can be obtained. 
We have also analyzed the response to weak magnetic field, and reproduced  the zigzag order in 2D Kitaev models by weakly coupling the chains on the honeycomb lattice. 
The correct response behavior and the magnetic pattern of zigzag order can only be obtained by applying the nonsymmorphic bosonization formulas. 
The nonsymmorphic bosonization method presented in this work is generalizable to other 1D nonsymmorphic systems including Kitaev spin ladders and doped 1D Kitaev systems,
which can be useful to gain deeper understanding and reveal more delicate physical properties in those systems.

\section*{METHODS}

\subsection*{DMRG numerics\\}

In all DMRG simulations in this work,
 the bond dimension $m$ and truncation error $\epsilon$ are taken as $m=1400$ and $\epsilon=10^{-9}$.
The system is chosen as containing $L=144$ sites with periodic boundary conditions. 

\subsection*{Bosonization\\}

In the Luttinger liquid theory,
the scaling dimensions for $J^x,J^y,J^z,N^x,N^y,N^z$ are $\kappa+1/(4\kappa)$,  $\kappa+1/(4\kappa)$, $1$, $1/(4\kappa)$, $1/(4\kappa)$, and $\kappa$, respectively \cite{Giamarchi2004}. 
Using the scaling dimensions of $J^\alpha,N^\alpha$ ($\alpha=x,y,z$), we obtain
\begin{eqnarray}
\langle J^x(0)J^x(r)\rangle&=&\langle J^y(0)J^y(r)\rangle=-\frac{1}{r^{2\kappa+\frac{1}{2\kappa}}},\nonumber\\
\langle J^z(0)J^z(r)\rangle&=&-\frac{1}{r^2},\nonumber\\
\langle N^x(0)N^x(r)\rangle &=&\langle N^y(0)N^y(r) \rangle=\frac{1}{r^{\frac{1}{2\kappa}}},\nonumber\\
\langle N^z(0)N^z(r)\rangle&=&\frac{1}{r^{2\kappa}},
\label{eq:JN_corr}
\end{eqnarray}
in which the UV cutoff $a$ is set to $1$.
The normalizations of $J^\alpha$ and $N^\alpha$ are taken such that the coefficients in the correlation functions in Eq. (\ref{eq:JN_corr}) are normalized to $1$.
Notice that actual value of $a$ is not important, since it can be absorbed into a redefinition of the coefficients in the bosonization formulas.
The expressions of the spin correlation functions  in  the main text are calculated with Eq. (\ref{eq:JN_corr}), by applying the nonsymmorphic bosonization formulas in Eq. (\ref{eq:bosonization_fourier}).

On the SU(2)$_1$ line in Fig. \ref{fig:phase} (a),
the low energy field theory is the  SU(2)$_1$ Wess-Zumino-Witten (WZW) model \cite{Affleck1988}.
The SU(2)$_1$ WZW model is defined by the following Sugawara Hamiltonian 
\begin{eqnarray}
\mathcal{H}=\frac{2\pi}{3}v\int dx (\vec{J}_L\cdot \vec{J}_L+\vec{J}_R\cdot \vec{J}_R)-\lambda\int dx \vec{J}_L\cdot \vec{J}_R,
\label{eq:Low_ham_SU2}
\end{eqnarray}
in which $\vec{J}_\alpha$ ($\alpha=L,R$) are WZW current operators forming Kac-Moody algebras, $v$ is the velocity, and $\lambda>0$ is the coupling of the marginal operator $\vec{J}_L\cdot \vec{J}_R$.
The $\vec{J}_L\cdot \vec{J}_R$ term introduces logarithmic corrections to the physical quantities.
For example, the staggered component of the spin-spin correlation functions are predicted by Eq. (\ref{eq:Low_ham_SU2}) to be
\begin{eqnarray}
S^{\alpha\alpha}_\pi\sim\frac{\ln^{1/2}(r/r_0)}{r},
\label{eq:ln_correction}
\end{eqnarray}
in which $\alpha=x,y,z$; $r$ should be replaced by $r_L=\frac{L}{\pi} \sin(\frac{\pi r}{L})$ on a finite size system;
$r_0$ is some short distance cutoff.

\section*{DATA AVAILABILITY}
\noindent The data that support the findings of this study are available from the authors upon reasonable request.


\section*{ACKNOWLEDGEMENTS}
\noindent 
W.Y. and I.A. acknowledge support from NSERC Discovery Grant 04033-2016.
C.X. is partially supported by Strategic Priority Research Program of CAS (No. XDB28000000).
A.N. acknowledges computational resources and services provided by Compute Canada and Advanced Research Computing at the University of
British Columbia. A.N. acknowledges support from the Max Planck-UBC-UTokyo Center for Quantum Materials and the Canada First Research Excellence Fund
(CFREF) Quantum Materials and Future Technologies Program of the Stewart Blusson Quantum Matter Institute (SBQMI).

\section*{COMPETING INTERESTS}
\noindent The authors declare no competing interests.

\section*{AUTHOR CONTRIBUTIONS}
\noindent 

W.Y. did the analytical analysis.
C.X. performed the numerical simulations.
S.X. and A.N. contributed to the numerical simulations.
I.A. supervised the project.

\section*{ADDITIONAL INFORMATION}
\noindent \textbf{Supplementary information} is available in the online version of the paper.\\



\end{document}


\title{Supplementary Materials: 
Nonsymmorphic bosonization in one-dimensional generalized Kitaev spin-1/2 models
}

\author{Wang Yang}
\thanks{These two authors contributed equally to this work.}
\affiliation{Department of Physics and Astronomy and Stewart Blusson Quantum Matter Institute,
University of British Columbia, Vancouver, B.C., Canada, V6T 1Z1}

\author{Chao Xu}
\thanks{These two authors contributed equally to this work.}
\affiliation{Kavli Institute for Theoretical Sciences, University of Chinese Academy of Sciences, Beijing 100190, China}

\author{Shenglong Xu}
\affiliation{Department of Physics \& Astronomy, Texas A\&M University, College Station, Texas, 77843, U.S.A.}

\author{Alberto Nocera}
\affiliation{Department of Physics and Astronomy and Stewart Blusson Quantum Matter Institute, 
University of British Columbia, Vancouver, B.C., Canada, V6T 1Z1}

\author{Ian Affleck}
\affiliation{Department of Physics and Astronomy and Stewart Blusson Quantum Matter Institute, 
University of British Columbia, Vancouver, B.C., Canada, V6T 1Z1}

\maketitle

\tableofcontents

\section{Hamiltonians in the original and rotated frames}
\label{app:Ham}

In this section, for the convenience of readers, we give the explicit expressions of the Hamiltonians before and after the four-sublattice rotations as well as the form of the four-sublattice rotation. 
For simplification of notation, we write the Hamiltonian $H$ as $H=\sum_{j=1}^L H_{j,j+1}$ where $H_{j,j+1}$ is the term on the bond between the sites $j$ and $j+1$.   

In the unrotated frame, $H_{j,j+1}$ is two-site periodic, which has the form 
\begin{eqnarray}
H_{2n+1,2n+2}&=&K S_{2n+1}^x S_{2n+2}^x +\Gamma (S_{2n+1}^y S_{2n+2}^z+S_{2n+1}^z S_{2n+2}^y)+J\vec{S}_{2n+1}\cdot \vec{S}_{2n+2}, \nn\\
H_{2n+2,2n+3}&=&K S_{2n+2}^y S_{2n+3}^y +\Gamma (S_{2n+2}^z S_{2n+3}^x+S_{2n+2}^x S_{2n+3}^z)+J\vec{S}_{2n+2}\cdot \vec{S}_{2n+3}.
\end{eqnarray}
The four-sublattice rotation $U_4$ is defined as
\begin{eqnarray}
\text{Sublattice $1$}: & (x,y,z) & \rightarrow (-x^{\prime},y^{\prime},-z^{\prime}),\nn\\ 
\text{Sublattice $2$}: & (x,y,z) & \rightarrow (-x^{\prime},-y^{\prime},z^{\prime}),\nn\\
\text{Sublattice $3$}: & (x,y,z) & \rightarrow (x^{\prime},-y^{\prime},-z^{\prime}),\nn\\
\text{Sublattice $4$}: & (x,y,z) & \rightarrow (x^{\prime},y^{\prime},z^{\prime}),
\label{eq:4rotation}
\end{eqnarray}
in which ``Sublattice $i$" ($1\leq i \leq 4$) represents all the sites $i+4n$ ($n\in \mathbb{Z}$) in the chain, and we have again dropped the spin symbol $S$ for simplicity.
After the four-sublattice rotation, the Hamiltonian becomes
\begin{eqnarray}
H^{\prime\prime}_{4n+1,4n+2}&=& (K+2J)S_{4n+1}^xS_{4n+2}^x-J\vec{S}_{4n+1}\cdot\vec{S}_{4n+2}+\Gamma (S^y_{4n+1}S^z_{4n+2}+S^z_{4n+1}S^y_{4n+2}), \nn\\
H^{\prime\prime}_{4n+2,4n+3}&=& (K+2J)S_{4n+2}^yS_{4n+3}^y-J\vec{S}_{4n+2}\cdot\vec{S}_{4n+3}+\Gamma (S^z_{4n+2}S^x_{4n+3}+S^x_{4n+2}S^z_{4n+3}), \nn\\
H^{\prime\prime}_{4n+3,4n+4}&=& (K+2J)S_{4n+3}^xS_{4n+4}^x-J\vec{S}_{4n+3}\cdot\vec{S}_{4n+4}-\Gamma (S^y_{4n+3}S^z_{4n+4}+S^z_{4n+3}S^y_{4n+4}), \nn\\
H^{\prime\prime}_{4n+4,4n+5}&=& (K+2J)S_{4n+4}^yS_{4n+5}^y-J\vec{S}_{4n+4}\cdot\vec{S}_{4n+5}-\Gamma (S^z_{4n+4}S^x_{4n+5}+S^x_{4n+4}S^z_{4n+5}).
\end{eqnarray}

\section{The nonsymmorphic symmetries}

\subsection{The nonsymmorphic symmetry group}
\label{app:sym_group}

\begin{figure}[h]
\includegraphics[width=8cm]{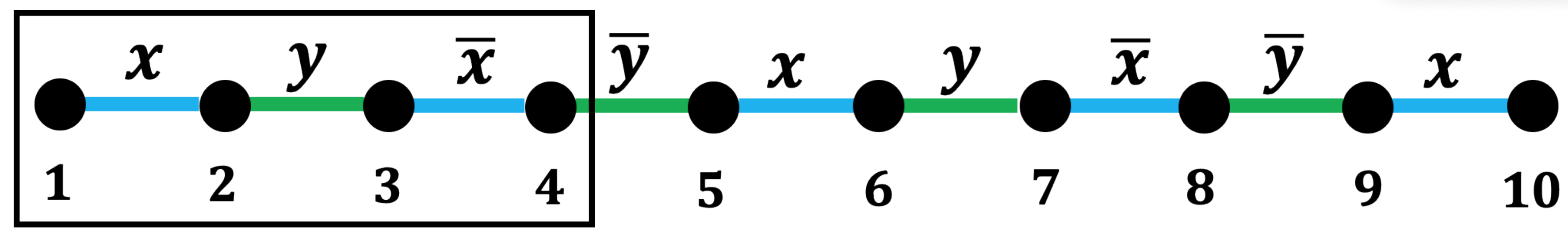}
\caption{The bond patterns of the Kitaev-Heisenberg-Gamma chain  after the four-sublattice rotation.
} \label{fig:bond4}
\end{figure}

In the four-sublattice rotated frame, the bond pattern in Fig. \ref{fig:bond4} is invariant under the following symmetry operations:
\begin{eqnarray}
T &: & (S_i^x,S_i^y,S_i^z)\rightarrow (-S_{i}^x,-S_{i}^y,-S_{i}^z)\nn\\
R(\hat{y},\pi)I&: & (S_i^x,S_i^y,S_i^z)\rightarrow (-S_{5-i}^x,S_{5-i}^y,-S_{5-i}^z)\nn\\
R(\hat{z},-\frac{\pi}{2})T_a&:& (S_i^x,S_i^y,S_i^z)\rightarrow (-S_{i+1}^y,S_{i+1}^x,S_{i+1}^z),
\label{eq:symmetries}
\end{eqnarray}
in which periodic boundary condition (i.e., $\vec{S}_{L+1}=\vec{S}_{1}$ where $L$ is the system size) is assumed, 
and the symmetry group $G$ is
\bea
G=\mathopen{<} T,R(\hat{y},\pi)I, R(\hat{z},-\frac{\pi}{2})T_a\mathclose{>}.
\eea
Since $T_{4a}=(R(\hat{z},-\frac{\pi}{2})T_a)^4$ is a symmetry and $\mathopen{<}T_{4a}\mathclose{>}$ is an abelian normal subgroup of $G$, we have the short exact sequence,
\bea
1\rightarrow \mathopen{<}T_{4a}\mathclose{>}\rightarrow G\rightarrow G/\mathopen{<}T_{4a}\mathclose{>} \rightarrow 1.
\eea

It has been proved in Ref. \onlinecite{Yang2020a} that $G/\mathopen{<}T_{4a}\mathclose{>}$ is isomorphic with $D_{4d}$.
Here we briefly review the proof. 
There is a generator-relation representation for the $D_n$ group as 
\begin{eqnarray}
D_n=\mathopen{<} \alpha,\beta| \alpha^n=\beta^2=(\alpha\beta)^2=e \mathclose{>}.
\label{eq:generator_Dn}
\end{eqnarray}
By defining $a=R(\hat{z},-\frac{\pi}{2})T_a$, $b=R(\hat{y},\pi)I$, 
it is straightforward to verify that  $a^4=T_{4a}$, $b^2=(ab)^2=1$,
hence Eq. (\ref{eq:generator_Dn}) is satisfied in the sense of modulo $T_{4a}$.
This implies that $G/\mathopen{<}T,T_{4a}\mathclose{>}$ is a subgroup of $D_{4}$.
To show that they are actually equal, it is enough to prove that the number of group elements in $G/\mathopen{<}T,T_{4a}\mathclose{>}$ is no less than that of $D_{4}$. 
In fact, since the actions of 
\begin{eqnarray}
\{1,a,a^2,a^3,b,ab,a^2b,a^3b\}
\end{eqnarray} 
restricted within the spin space are 
\begin{eqnarray}
&\{1,R(\hat{z},\frac{\pi}{2}),R(\hat{z},\pi),R(\hat{z},-\frac{\pi}{2}),R(\hat{y},\pi),R(\frac{1}{\sqrt{2}}(\hat{x}-\hat{y}),\pi),R(\hat{x},\pi),R(\frac{1}{\sqrt{2}}(\hat{x}+\hat{y}),\pi)\},
\end{eqnarray}
which are all distinct operations,
there must be at least eight ($=|D_4|$) distinct group  elements in $G/\mathopen{<}T,T_{4a}\mathclose{>}$.
Hence we have $G/\mathopen{<}T,T_{4a}\mathclose{>}\cong D_4$, i.e., $G/\mathopen{<}T_{4a}\mathclose{>}\cong D_4\times \mathbb{Z}_2^T\cong D_{4d}$, where $\mathbb{Z}_2^T$ is the $\mathbb{Z}_2$ group generated by the time reversal operation. 
  
Next we show that $G$ is nonsymmorphic, i.e., within the following short exact sequence,
\bea
1\rightarrow \mathopen{<}T_{4a}\mathclose{>}\rightarrow G\xrightarrow[]{\pi} D_{4d}\rightarrow 1
\eea
there does not exist a group homomorphism $\tau:D_{4d}\rightarrow G$, such that $\pi\tau$ is the identity map on $D_{4d}$.
We can prove this by contradiction. 
Suppose  that such a $\tau$ exists. 
Let $y=\pi(R(\hat{z},-\frac{\pi}{2})T_{a})$.
Since $\pi(\tau(y))=\pi(R(\hat{z},-\frac{\pi}{2})T_{a})=y$,  $\tau(y)$ and $R(\hat{z},-\frac{\pi}{2})T_{a}$ must only differ by an element in $\text{Ker}(\pi)=\mathopen{<} T_{4a}\mathclose{>}$, i.e., there exists $n\in \mathbb{Z}$, such that $\tau(y)=R(\hat{z},-\frac{\pi}{2})T_{(1+4n)a}$.
Since $\tau$ is assumed to be a group homomorphism, we have
\bea
\tau(y^4)=[\tau(y)]^4=T_{(1+4n)4a}.
\label{eq:y3}
\eea
Notice that $T_{(1+4n)4a}\in \text{Ker}(\pi)$, hence
\bea
\pi(\tau(y^4))=e,
\eea
where $e$ is the identity element in $D_{4d}$.
On the other hand, recall that  by assumption $\pi\cdot \tau=1$, therefore $y^4=e$.
However, for $\tau$ to be a group homomorphism, we must have
\bea
\tau(y^4)=\tau(e)=e_0,
\label{eq:y3_b}
\eea
where $e_0$ is the identity element in $G$.
It is clear that  Eq. (\ref{eq:y3_b}) contradicts with Eq. (\ref{eq:y3}), since $1+4n$ can never be $0$ (so that $T_{(1+4n)4a}\neq e_0$ for whatever $n\in \mathbb{Z}$).

\subsection{Symmetry transformation properties of the bosonized fields}
\label{app:sym_bosonized_field}

The symmetry transformation properties of the bosonized fields $\theta$ and $\varphi$ are
\bea
T_a&:& \theta\rightarrow \theta +\sqrt{\pi},~\varphi\rightarrow \varphi+\frac{\sqrt{\pi}}{2},\nn\\
R(\hat{y},\pi)&:&\theta\rightarrow -\theta +\sqrt{\pi}, ~\varphi\rightarrow -\varphi,\nn\\
R(\hat{z},\beta)&:& \theta\rightarrow \theta+\frac{\beta}{\sqrt{\pi}},~\varphi\rightarrow \varphi,
\eea
in which the actions are local in space and time, and
\bea
T &:& \theta(t,x)\rightarrow \theta(-t,x)+\sqrt{\pi},~ \varphi(t,x)\rightarrow -\varphi(-t,x),\nn\\
I &:& \theta(t,x)\rightarrow \theta(t,-x)+\sqrt{\pi},~\varphi(t,x)\rightarrow \varphi(t,-x)+\frac{\sqrt{\pi}}{2}.
\eea

It can be proved that with these transformations, we have
\bea
T_a (J^x,J^y,J^z)(T_a)^{-1}&=&(J^x,J^y,J^z),\nn\\
T_a (N^x,N^y,N^z)(T_a)^{-1}&=&(-N^x,-N^y,-N^z),
\eea
\bea
R(\hat{y},\pi)(J^x,J^y,J^z)[R(\hat{y},\pi)]^{-1}&=&(-J^x,J^y,-J^z),\nn\\
R(\hat{y},\pi)(N^x,N^y,N^z)[R(\hat{y},\pi)]^{-1}&=&(-N^x,N^y,-N^z),
\eea
\bea
R(\hat{z},\beta) (J^x,J^y,J^z)  [R(\hat{z},\beta)]^{-1} &=& (\cos(\beta)J^x+\sin(\beta)J^y,-\sin(\beta)J^x+\cos(\beta)J^y,J^z),\nn\\
R(\hat{z},\beta) (N^x,N^y,N^z)  [R(\hat{z},\beta)]^{-1} &=& (\cos(\beta)N^x+\sin(\beta)N^y,-\sin(\beta)N^x+\cos(\beta)N^y,N^z),
\eea
\bea
T(J^x(t,x),J^y(t,x),J^z(t,x))T^{-1}&=&(-J^x(-t,x),-J^y(-t,x),-J^z(-t,x)),\nn\\
T(N^x(t,x),N^y(t,x),N^z(t,x))T^{-1}&=&(-N^x(-t,x),-N^y(-t,x),-N^z(-t,x)),
\eea
\bea
I(J^x(t,x),J^y(t,x),J^z(t,x))I^{-1}&=&(J^x(-t,x),J^y(-t,x),J^z(-t,x)),\nn\\
I(N^x(t,x),N^y(t,x),N^z(t,x))I^{-1}&=&(-N^x(t,-x),-N^y(t,-x),-N^z(t,-x)).
\eea

\section{The nonsymmorphic bosonization formulas}
\subsection{Derivations of the nonsymmorphic bosonization formulas}
\label{app:nonsym_bosonization}

The formula $S^{\alpha}_{i+4n} = D^{(i)}_{\alpha\beta} J^{\beta}(x)+(-)^i C^{(i)}_{\alpha\beta} N^{\beta}(x)$ in the main text can alternatively be written as
\begin{flalign}
(S^{x}_{i+4n}~S^{ y}_{i+4n}~S^{ z}_{i+4n})&=
(J^{ x}~J^{ y}~J^{ z})D^{(i)}
+(-)^{i}(N^{ x}~N^{ y}~N^{ z})C^{(i)},
\label{eq:abelian_LL1_matrix}
\end{flalign}
in which $D^{(i)},C^{(i)}$ ($i=1,2,3,4$) are $3\times 3$ matrices.
Recall that the symmetry group in the four-sublattice rotated frame is
\bea
G_3=\mathopen{<} T,R(\hat{y},\pi)T_a I,R(\hat{z},-\frac{\pi}{2})T_a \mathclose{>}.
\label{eq:generate_G3}
\eea
The generators $R(\hat{z},-\pi/2)T_a$ and $R(\hat{y},\pi)T_aI$ require
\bea
C^{(2)}=M_z C^{(1)} (M_z)^{-1},~C^{(3)}=M_zC^{(2)} (M_z)^{-1},~
C^{(4)}=M_zC^{(3)} (M_z)^{-1},~C^{(1)}=M_zC^{(4)} (M_z)^{-1},
\label{eq:relations_Cprime}
\eea
and
\bea
C^{(4)}=M_yC^{(1)} (M_y)^{-1},~
C^{(4)}=M_yC^{(2)} (M_y)^{-1},
\label{eq:relations_Cprime_2}
\eea
in which 
\bea
M_z=\left(
\begin{array}{ccc}
0&1&0\\
-1&0&0\\
0&0&1
\end{array}
\right),~
M_y=\left(
\begin{array}{ccc}
-1&0&0\\
0&1&0\\
0&0&-1
\end{array}
\right).
\eea
The requirements in Eq. (\ref{eq:relations_Cprime_2}) lead to the following single constraint on $C^{(1)}$,
\bea
C^{(1)}&=&M_{xy}C^{(1)} M_{xy},
\label{eq:eq_C1prime}
\eea
in which 
\bea
M_{xy}=\left(
\begin{array}{ccc}
0&1&0\\
1&0&0\\
0&0&-1
\end{array}
\right).
\eea
The solution of Eq. (\ref{eq:eq_C1prime}) is
\bea
C^{(1)}=\left(
\begin{array}{ccc}
a_C& b_C&c_C\\
b_C&a_C&-c_C\\
h_C&-h_C&i_C
\end{array}
\right).
\eea
The other matrices $C^{(i)}$ ($i=2,3,4$) can be obtained from Eq. (\ref{eq:relations_Cprime}).
The treatments for the $D$-coefficients are similar. 

In the four-sublattice rotated frame, the explicit forms of the  nonsymmorphic bosonization formulas are given by
\bea
S_{1+4n}^x&=&a_DJ^x+b_DJ^y+h_DJ^z-a_CN^x-b_CN^y-h_CN^z,\nn\\
S_{1+4n}^y&=&b_DJ^x+a_DJ^y-h_DJ^z-b_CN^x-a_CN^y+h_CN^z,\nn\\
S_{1+4n}^z&=&c_DJ^x-c_DJ^y+i_DJ^z-c_CN^x+c_CN^y-i_C N^z,
\label{eq:bosonization_S1}
\eea
\bea
S_{2+4n}^x&=&a_DJ^x-b_DJ^y-h_DJ^z+a_CN^x-b_CN^y-h_CN^z,\nn\\
S_{2+4n}^y&=&-b_DJ^x+a_DJ^y-h_DJ^z-b_CN^x+a_CN^y-h_CN^z,\nn\\
S_{2+4n}^z&=&-c_DJ^x-c_DJ^y+i_DJ^z-c_CN^x-c_CN^y+i_C N^z,
\label{eq:bosonization_S2}
\eea
\bea
S_{3+4n}^x&=&a_DJ^x+b_DJ^y-h_DJ^z-a_CN^x-b_CN^y+h_CN^z,\nn\\
S_{3+4n}^y&=&b_DJ^x+a_DJ^y+h_DJ^z-b_CN^x-a_CN^y-h_CN^z,\nn\\
S_{3+4n}^z&=&-c_DJ^x+c_DJ^y+i_DJ^z+c_CN^x-c_CN^y-i_C N^z,
\label{eq:bosonization_S3}
\eea
\bea
S_{4+4n}^x&=&a_DJ^x-b_DJ^y+h_DJ^z+a_CN^x-b_CN^y+h_CN^z,\nn\\
S_{4+4n}^y&=&-b_DJ^x+a_DJ^y+h_DJ^z-b_CN^x+a_CN^y+h_CN^z,\nn\\
S_{4+4n}^z&=&c_DJ^x+c_DJ^y+i_DJ^z+c_CN^x+c_CN^y+i_C N^z.
\label{eq:bosonization_S4}
\eea
The above nonsymmorphic bosonization  formulas can be written in terms of the Fourier modes as
\bea
S_j^\alpha&=&(a_DJ^\alpha-b_CN^{\beta})+(-)^j(a_CN^{\alpha}-b_DJ^{\beta})+\sqrt{2}\cos\big(\frac{\pi}{2}(j+\frac{\eta_\alpha}{2})\big)h_CN^z+\sqrt{2}\eta_\alpha\sin\big(\frac{\pi}{2}(j+\frac{\eta_\alpha}{2})\big)h_DJ^z,\nn\\
S_j^z&=&i_DJ^z+\sqrt{2}\cos\big(\frac{\pi}{2}(j+\frac{1}{2})\big) (c_DJ^y+c_CN^x)+(-)^ji_CN^z+\sqrt{2}\sin\big(\frac{\pi}{2}(j+\frac{1}{2})\big) (c_DJ^x+c_CN^y),
\label{eq:bosonization_fourier_b}
\eea
in which $\alpha,\beta\in \{x,y\}$, $\alpha\neq \beta$,  and $\eta_x=-\eta_y=1$.

In the original frame without sublattice rotation, the bosonization formulas are 
\bea
S_{1+4n}^x&=&-a_DJ^x-b_DJ^y-h_DJ^z+a_CN^x+b_CN^y+h_CN^z,\nn\\
S_{1+4n}^y&=&b_DJ^x+a_DJ^y-h_DJ^z-b_CN^x-a_CN^y+h_CN^z,\nn\\
S_{1+4n}^z&=&-c_DJ^x+c_DJ^y-i_DJ^z+c_CN^x-c_CN^y+i_C N^z,
\label{eq:bosonization_S1_b}
\eea
\bea
S_{2+4n}^x&=&-a_DJ^x+b_DJ^y+h_DJ^z-a_CN^x+b_CN^y+h_CN^z,\nn\\
S_{2+4n}^y&=&b_DJ^x-a_DJ^y+h_DJ^z+b_CN^x-a_CN^y+h_CN^z,\nn\\
S_{2+4n}^z&=&-c_DJ^x-c_DJ^y+i_DJ^z-c_CN^x-c_CN^y+i_C N^z,
\label{eq:bosonization_S2_b}
\eea
\bea
S_{3+4n}^x&=&a_DJ^x+b_DJ^y-h_DJ^z-a_CN^x-b_CN^y+h_CN^z,\nn\\
S_{3+4n}^y&=&-b_DJ^x-a_DJ^y-h_DJ^z+b_CN^x+a_CN^y+h_CN^z,\nn\\
S_{3+4n}^z&=&c_DJ^x-c_DJ^y-i_DJ^z-c_CN^x+c_CN^y+i_C N^z,
\label{eq:bosonization_S3_b}
\eea
\bea
S_{4+4n}^x&=&a_DJ^x-b_DJ^y+h_DJ^z+a_CN^x-b_CN^y+h_CN^z,\nn\\
S_{4+4n}^y&=&-b_DJ^x+a_DJ^y+h_DJ^z-b_CN^x+a_CN^y+h_CN^z,\nn\\
S_{4+4n}^z&=&c_DJ^x+c_DJ^y+i_DJ^z+c_CN^x+c_CN^y+i_C N^z.
\label{eq:bosonization_S4_b}
\eea
From Eqs. (\ref{eq:bosonization_S1_b},\ref{eq:bosonization_S2_b},\ref{eq:bosonization_S3_b},\ref{eq:bosonization_S4_b}), it is straightforward to verify that the sums of the spins within a unit cell are
\bea
\sum_{i=1}^4 S_{i+4n}^x=4h_CN^z,~\sum_{i=1}^4 S_{i+4n}^y=4h_CN^z,~ \sum_{i=1}^4 S_{i+4n}^z=4i_CN^z.
\label{eq:low_energy_field_term}
\eea
When a nonzero expectation value of $N^z$ is developed, the spin magnetization is $\langle N^z \rangle(h_C,h_C,i_C)$ on all sites.
This shows that the response of the spin magnetization is fixed to the $(h_C,h_C,i_C)$-direction, regardless of the direction of the applied weak magnetic field. 

\subsection{Spin correlation functions}

According to Eqs. (\ref{eq:bosonization_S1},\ref{eq:bosonization_S2},\ref{eq:bosonization_S3},\ref{eq:bosonization_S4}),
within the four-sublattice rotated frame,
the equal-time spin-spin correlation functions can be calculated as
\bea
\begin{array}{ccccccc}
 & \frac{a^2_D}{r^{2\kappa+1/(2\kappa)}} & \frac{b_D^2}{r^{2\kappa+1/(2\kappa)}} & \frac{h_D^2}{r^2} & \frac{a_C^2}{r^{1/(2\kappa)}} & \frac{b_C^2}{r^{1/(2\kappa)}} & \frac{h_C^2}{r^{2\kappa}} \\
 \left<S_1^xS_{1+4n}^x\right>& -1&-1&-1&1&1&1\\
 \left<S_1^xS_{2+4n}^x\right>& -1& 1& 1&-1&1&1\\
 \left<S_1^xS_{3+4n}^x\right>& -1&-1& 1&1&1&-1\\
 \left<S_1^xS_{4+4n}^x\right>& -1& 1&-1&-1&1&-1\\
\end{array}
\label{eq:pattern_x}
\eea
\bea
\begin{array}{ccccccc}
 & \frac{a^2_D}{r^{2\kappa+1/(2\kappa)}} & \frac{b_D^2}{r^{2\kappa+1/(2\kappa)}} & \frac{h_D^2}{r^2} & \frac{a_C^2}{r^{1/(2\kappa)}} & \frac{b_C^2}{r^{1/(2\kappa)}} & \frac{h_C^2}{r^{2\kappa}} \\
 \left<S_1^yS_{1+4n}^y\right>&-1&-1&-1&1&1&1\\
 \left<S_1^yS_{2+4n}^y\right>&-1& 1&-1&-1&1&-1\\
 \left<S_1^yS_{3+4n}^y\right>&-1&-1& 1&1&1&-1\\
 \left<S_1^yS_{4+4n}^y\right>&-1& 1& 1&-1&1&1\\
\end{array}
\label{eq:pattern_yy}
\eea
\bea
\begin{array}{ccccccc}
 & \frac{c^2_D}{r^{2\kappa+1/(2\kappa)}} & \frac{c_D^2}{r^{2\kappa+1/(2\kappa)}} & \frac{i_D^2}{r^2} & \frac{c_C^2}{r^{1/(2\kappa)}} & \frac{c_C^2}{r^{1/(2\kappa)}} & \frac{i_C^2}{r^{2\kappa}} \\
 \left<S_z^zS_{1+4n}^z\right>& -1&-1&-1&1&1&1\\
 \left<S_1^zS_{2+4n}^z\right>&  1&-1&-1&1&-1&-1\\
 \left<S_1^zS_{3+4n}^z\right>&  1& 1&-1&-1&-1&1\\
 \left<S_1^zS_{4+4n}^z\right>& -1& 1&-1&-1&1&-1\\
\end{array}.
\label{eq:pattern_zz}
\eea
We note that the difference between the $xx$ and $yy$ correlations has a simple form
\bea
\begin{array}{ccccccc}
 & \frac{a^2_D}{r^{2\kappa+1/(2\kappa)}} & \frac{b_D^2}{r^{2\kappa+1/(2\kappa)}} & \frac{h_D^2}{r^2} & \frac{a_C^2}{r^{1/(2\kappa)}} & \frac{b_C^2}{r^{1/(2\kappa)}} & \frac{h_C^2}{r^{2\kappa}} \\
 \left<S_1^xS_{1+4n}^x-S_1^yS_{1+4n}^y\right>& 0 & 0 & 0 & 0 & 0 & 0\\
 \left<S_1^xS_{2+4n}^x-S_1^yS_{2+4n}^y\right>& 0 & 0 & 2 & 0 & 0 & 2\\
 \left<S_1^xS_{3+4n}^x-S_1^yS_{3+4n}^y\right>& 0 & 0 & 0 & 0 & 0 & 0\\
 \left<S_1^xS_{4+4n}^x-S_1^yS_{4+4n}^y\right>& 0 & 0 &-2 & 0 & 0 &-2\\
\end{array}.
\label{eq:pattern_diff}
\eea

\subsection{Method for determining  the signs of the bosonization coefficients}
\label{app:determine_sign}

In this section, we discuss the method to numerically determine the signs of the ten bosonization coefficients. 
We will work in the four-sublattice rotated frame. 

\subsubsection{The positive coefficients $a_D,a_C,i_D,i_C$}

The coefficients will be divided into several groups according to their corresponding Fourier wavevectors.
We will see that only the relative signs of the coefficients within each group can be determined by numerics.
However, we know that at least for small (possibly also moderate) values of  $K^\prime=K+2J$ and $\Gamma$,
the coefficients $a_D,a_C,i_D,i_C$ are positive, and all other coefficients are small (since they vanish when $K^\prime=\Gamma=0$).
Therefore, we will assume $a_D>0$, $a_C>0$, $i_D>0$, $i_C>0$,
and use various responses  or correlation functions to determine the signs of the other six coefficients.

\subsubsection{Determining sign of $i_C$}

Consider adding a small staggered magnetic field along the $z$-direction, i.e.,
\bea
-h^z_\pi \sum_n (-S_{1+4n}^z+S_{2+4n}^z-S_{3+4n}^z+S_{4+4n}^z).
\label{eq:h_zpi}
\eea
According to Eqs. (\ref{eq:bosonization_S1},\ref{eq:bosonization_S2},\ref{eq:bosonization_S3},\ref{eq:bosonization_S4}),
the low energy Hamiltonian corresponding to Eq. (\ref{eq:h_zpi}) is 
\bea
-h^z_\pi i_C \int dx N^z.
\eea
Since this is a relevant perturbation, a gap opens and $\langle N^z \rangle\neq 0$.
As a result, the spin expectation values are given by Eqs. (\ref{eq:bosonization_S1},\ref{eq:bosonization_S2},\ref{eq:bosonization_S3},\ref{eq:bosonization_S4}) to be 
\bea
\langle \vec{S}_{1+4n} \rangle &=& \langle N^z \rangle(-h_C,h_C,-i_C),\nn\\
\langle \vec{S}_{2+4n} \rangle &=& \langle N^z \rangle(-h_C,-h_C,i_C),\nn\\
\langle \vec{S}_{3+4n} \rangle &=& \langle N^z \rangle(h_C,-h_C,-i_C),\nn\\
\langle \vec{S}_{4+4n} \rangle &=& \langle N^z \rangle(h_C,h_C,i_C).
\eea
This clearly gives the ratio $h_C/i_C$.
By assuming the sign of $i_C$, the sign of $h_C$ is determined.

\subsubsection{Determining signs of $b_C$ and $c_C$}

Next we add a small staggered field $h^x_\pi$ along the $x$-direction, i.e., 
\bea
-h_\pi^x \sum_n (-S_{1+4n}^x+S_{2+4n}^x-S_{3+4n}^x+S_{4+4n}^x).
\eea
The low energy Hamiltonian is 
\bea
-h^x_\pi \big(a_C \int dx N^x-b_D\int dx J^y\big).
\eea
Since $N^x$ has a smaller scaling dimension than $J^y$,
we expect the $\int dx N^x$ term dominates over $\in dx J^y$.
Plugging $\langle N^x\rangle\neq 0$ into Eqs. (\ref{eq:bosonization_S1},\ref{eq:bosonization_S2},\ref{eq:bosonization_S3},\ref{eq:bosonization_S4}), we obtain
\bea
\langle \vec{S}_{1+4n} \rangle &=& \langle N^x \rangle(-a_C,-b_C,-c_C),\nn\\
\langle \vec{S}_{2+4n} \rangle &=& \langle N^x \rangle(a_C,-b_C,-c_C),\nn\\
\langle \vec{S}_{3+4n} \rangle &=& \langle N^x \rangle(-a_C,-b_C,c_C),\nn\\
\langle \vec{S}_{4+4n} \rangle &=& \langle N^x \rangle(a_C,-b_C,c_C).
\eea
Clearly this gives the ratios $b_C/a_C$ and $c_C/a_C$, from which the signs of $b_C$ and $c_C$ can be determined by assuming a positive $a_C$.

\subsubsection{Determining sign of $h_D$}

Consider adding a small uniform magnetic field along the $z$-direction, i.e.,
\bea
-h^z_0 \sum_n (S_{1+4n}^z+S_{2+4n}^z+S_{3+4n}^z+S_{4+4n}^z).
\label{eq:h_z0}
\eea
The low energy Hamiltonian  is 
\bea
-h^z_0 i_D \int dx J^z.
\eea
Plugging $\langle J^z\rangle\neq 0$ into Eqs. (\ref{eq:bosonization_S1},\ref{eq:bosonization_S2},\ref{eq:bosonization_S3},\ref{eq:bosonization_S4}), we obtain
\bea
\langle \vec{S}_{1+4n} \rangle &=& \langle J^z \rangle(h_D,-h_D,i_D),\nn\\
\langle \vec{S}_{2+4n} \rangle &=& \langle J^z \rangle(-h_D,-h_D,i_D),\nn\\
\langle \vec{S}_{3+4n} \rangle &=& \langle J^z \rangle(-h_D,h_D,i_D),\nn\\
\langle \vec{S}_{4+4n} \rangle &=& \langle J^z \rangle(h_D,h_D,i_D).
\eea
Clearly this gives the ratio $h_D/i_D$, which determines the sign of $h_D$ assuming $i_D>0$.

\subsubsection{Determining signs of $b_D$ and $c_D$}

The small field in this case is more complicated, which is chosen as
\bea
-h^{xy}\big[ a_C\sum_n (S_{1+4n}^y+S_{2+4n}^y+S_{3+4n}^y+S_{4+4n}^y)+b_C\sum_n (-S_{1+4n}^x+S_{2+4n}^x-S_{3+4n}^x+S_{4+4n}^x)  \big].
\eea
The low energy Hamiltonian  is 
\bea
-h^{xy} (a_Ca_D-b_Cb_D) \int dx J^y.
\eea
Plugging $\langle J^y\rangle\neq 0$ into Eqs. (\ref{eq:bosonization_S1},\ref{eq:bosonization_S2},\ref{eq:bosonization_S3},\ref{eq:bosonization_S4}), we obtain
\bea
\langle \vec{S}_{1+4n} \rangle &=& \langle J^y \rangle(b_D,a_D,-c_D),\nn\\
\langle \vec{S}_{2+4n} \rangle &=& \langle J^y \rangle(-b_D,a_D,-c_D),\nn\\
\langle \vec{S}_{3+4n} \rangle &=& \langle J^y \rangle(b_D,a_D,c_D),\nn\\
\langle \vec{S}_{4+4n} \rangle &=& \langle J^y \rangle(-b_D,a_D,c_D).
\eea
Clearly this gives the ratios $b_D/a_D$ and $c_D/a_D$, which determines the sign of $b_D$ and $c_D$ assuming $a_D>0$.

\section{Methods for numerical fittings}
\label{app:numerics_fit}

In this section, we show how we numerically extract different Fourier components from the spin correlation functions. By separating different components (e.g. components with different momenta) of the correlation functions, we can determine the bosonization coefficients. 
\subsection{$\langle S^z_1 S^z_r \rangle$}
We denote
\begin{equation}
    \langle S^z_1 S^z_{r} \rangle = ZZ^i_r =ZZ^i_{4j+i}\,,
\end{equation}
where $i$ can take 1, 2, 3, and 4. 
Then the interpolation is done for each given $i$, e.g. $\mathbb{ZZ}^i_r$ is now a smooth function satisfying $\mathbb{ZZ}^i_{4j+i} = ZZ^i_{4j+i}$, and for $r\neq 4j+1$, $\mathbb{ZZ}^{i}_r$ is obtained by interpolation. Numerically we use standard cubic spline interpolation method.

The zero  and $\pi$ momentum components are relatively easy to extract. 
The zero momentum part is extracted by
\begin{equation}
    S^{zz}_0 (r) = \frac{1}{4}\left( \mathbb{ZZ}^1_r +\mathbb{ZZ}^2_r +\mathbb{ZZ}^3_r +\mathbb{ZZ}^4_r\right) \,.
\end{equation}
As predicted by the bosonization formula Eq.~\eqref{eq:pattern_zz}, 
\begin{equation}
    S^{zz}_0(r) = - \frac{i_D^2}{r^2}\,,
\end{equation}
therefore we can check this by taking a linear regression between ${\rm{log}}(-S^{zz}_0(r_L))$ and ${\rm{log}}(r_L)$. When we take the numerics we only use 61 sites from the central part of the system, which are sites from 43 to 103. 
The linear regression gives 
\begin{equation}
    \log_{10}(-S^{zz}_0) \simeq -1.995\times \log_{10}(r_L) - 1.478\,.
\end{equation}
Then the calculated exponent is 1.995 which is very close to 2, and $|i_D|$ can be estimated as
\begin{equation}
    |i_D|  \simeq \sqrt{10^{-1.478}} \simeq 0.182\,.
\end{equation}

The $\pi$ momentum part is extracted by
\begin{equation}
    S^{zz}_{\pi}(r) = \frac{1}{4}\left( \mathbb{ZZ}^1_r -\mathbb{ZZ}^2_r +\mathbb{ZZ}^3_r -\mathbb{ZZ}^4_r\right) \,,
\end{equation}
which should behave as (see Eq.~\eqref{eq:pattern_zz})
\begin{equation}
    S^{zz}_{\pi}(r) = \frac{i_C^2}{r^{2\kappa}}\,.
\end{equation}
Again, we take linear regression between $\log (r_L)$ and $\log (S^{zz}_{\pi}(r_L))$, and get
\begin{equation}
    \log_{10} \left( S^{zz}_{\pi}(r_L) \right) \simeq -1.367 \times \log_{10} (r_L) - 0.8796\,.
\end{equation}
Correspondingly, $|i_C|$ and $\kappa$ are estimated as
\begin{equation}
    |i_C| \simeq \sqrt{10^{-0.8796}} \simeq 0.363;~~ \kappa\simeq \frac{1.367}{2} \simeq 0.6835\,. 
\end{equation}

The $\pi/2$ component is much smaller than the zero and $\pi$ components as we expected, and it can be extracted by
\begin{equation}
    S^{zz}_{\pi/2}(r)  = \frac{1}{2} \left(\mathbb{ZZ}^1_r - \mathbb{ZZ}^3_r  \right).
\end{equation}
Then this $\pi/2$ component should be described by
\begin{equation}
    S^{zz}_{\pi/2}(r)  = \frac{c_C^2}{r^{2\kappa}} - \frac{c_D^2}{r^{2\kappa+1/(2\kappa)}}\,,
\end{equation}
cf. Eq.~\eqref{eq:pattern_zz}, and we use the curve fitting tool (cftool implemented by MATLAB) to fit it by the function with form $a/r^{c+1/c} + b/r^c$. Since the magnitude of $S^{zz}_{\pi/2}(r)$ is too small, we do not take the fitting directly but we multiply the correlation by $10^{3}$ first. Eventually we get
\begin{equation}
    10^3 \times S^{zz}_{\pi/2}(r_L) \simeq \frac{-1.290}{r_L^{0.7299+1/0.7299}} + \frac{0.5972}{r_L^{0.7299}}.
\end{equation}
The corresponding
\begin{align}
    |c_C| &\simeq \sqrt{0.5972\times10^{-3}} \simeq 0.0244\\
    |c_D| &\simeq \sqrt{1.290\times 10^{-3}} \simeq 0.0359\\
    \kappa &\simeq \frac{1}{2\times 0.7299} \simeq 0.6850.
\end{align}

\subsection{$\langle S^x_1 S^x_r \rangle$ and $\langle S^y_1 S^y_r \rangle$}
Similar as $\langle S^z_1 S^z_r \rangle$, we can define $XX^i_{4j+i}$, and the interpolated functions $\mathbb{XX}^i_r$. Both correlation functions with zero and $\pi$ momenta should be symmetric from the center of the system, however it is not perfectly symmetric numerically due the finite size effects. Here we take the average of $\langle S^x_1 S^x_r \rangle$ and $\langle S^y_1 S^y_r \rangle$ to make them symmetric manually.
The zero or $\pi$ momentum component are calculated by
\begin{align}
    S^{xx}_{0/\pi}(r) &= \frac{1}{4}\left( \mathbb{XX}^1_r \pm\mathbb{XX}^2_r +\mathbb{XX}^3_r \pm\mathbb{XX}^4_r \right) \\
    S^{yy}_{0/\pi}(r) &= \frac{1}{4}\left( \mathbb{YY}^1_r \pm\mathbb{YY}^2_r +\mathbb{YY}^3_r \pm\mathbb{YY}^4_r \right)\,,
\end{align}
and the average one
\begin{equation}
    S^{xx+yy}_{0/\pi}(r) = \frac{1}{2} \left( S^{xx}_{0/\pi}(r) + S^{yy}_{0/\pi}(r) \right)\,.
\end{equation}
According to Eq.~\eqref{eq:pattern_x} and \eqref{eq:pattern_yy}, 
\begin{equation}
    S^{xx+yy}_0(r) = \frac{b_C^2}{r^{2\kappa}} - \frac{a_D^2}{r^{2\kappa+1/(2\kappa)}}\,,
\end{equation}
and the fitting result reads
\begin{equation}
    10^3 \times S^{xx+yy}_0(r_L) \simeq \frac{0.001068}{r_L^{0.7339}} - \frac{25.59}{r_L^{0.7339+1/0.7339}}\,.
\end{equation}
Therefore the estimated coefficients are
\begin{align}
    |b_C| &\simeq \sqrt{0.001068\times10^{-3}}\simeq0.00103\\
    |a_D| &\simeq \sqrt{25.95\times10^{-3}}\simeq 0.161\\
    \kappa &\simeq \frac{1}{2\times 0.7339} \simeq 0.6813
\end{align}
According to Eq.~\eqref{eq:pattern_x} and \eqref{eq:pattern_yy},
\begin{equation}
    S^{xx+yy}_{\pi}(r) = \frac{a_C^2}{r^{2\kappa}} - \frac{b_D^2}{r^{2\kappa + 1/(2\kappa)}}\,,
\end{equation}
however it is hard to obtain a reliable result of $|b_D|$, since $b_D^2$ should be much smaller than $a_C^2$ and $1/r^{2\kappa+1/(2\kappa)}$ also decays faster than $1/r^{2\kappa}$, so at long distance, the contribution from $b_D^2$ term is much smaller the $a_C^2$ term.
Therefore we only fit $S^{xx+yy}_{\pi}(r)$ by $a_C^2/r^{2\kappa}$, and the results are
\begin{align}
    |a_C| &\simeq 0.129\\
    \kappa &\simeq 0.6805\,.
\end{align}

The $\pi/2$ component is much smaller than the zero and $\pi$ components, so we need to be more careful. The asymmetries in $S^{xx}_{0/\pi}$ and $S^{yy}_{0/\pi}$ will affect the extraction of the $\pi/2$ component, since the magnitude of the asymmetries are comparable to the magnitude of the $\pi/2$ component. To deal with this problem we can remove all the zero and $\pi$ component from $S^{xx}$ and $S^{yy}$, such that only  $\pi/2$ components remain, e.g.
\begin{align}
    \tilde{XX}_{\pi/2}(r) &= \langle S^x_1 S^x_r \rangle - S^{xx}_0(r) + (-1)^r S^{xx}_{\pi}(r)\\
    \tilde{YY}_{\pi/2}(r) &= \langle S^y_1 S^y_r \rangle - S^{yy}_0(r) + (-1)^r S^{yy}_{\pi}(r)\,.
\end{align}
Then we calculate the difference between $\tilde{XX}$ and $\tilde{YY}$
\begin{equation}
    \tilde{XY}_r = \tilde{XX}_{\pi/2}(r) - \tilde{YY}_{\pi/2}(r)\,.
\end{equation}
Eq.~\eqref{eq:pattern_diff} shows the result of difference between $\langle S^x_1 S^x_r \rangle$ and $\langle S^y_1 S^y_r \rangle$, where most of the terms are cancelled, while only $h_D^2$ and $h_C^2$ terms are left. Again, the interpolation of $\tilde{XY}_r$, $\mathbb{XY}^{i}_r$, can be calculated, and we only need the result of $i=2$, such that we can define $S^{xx-yy}_{\pi/2}(r) = 0.5\times\mathbb{XY}^2_r$, and it should be fitted by
\begin{equation}
    S^{xx-yy}_{\pi/2}(r) = \frac{h_C^2}{r^{2\kappa}} + \frac{h_D^2}{r^2}\,.
\end{equation}
The fitting result is 
\begin{equation}
    10^3 \times S^{xx-yy}_{\pi/2}(r_L) \simeq \frac{0.1893}{r_L^{1.363}} + \frac{0.7053}{r_L^2}\,,
\end{equation}
so the estimated coefficients are
\begin{align}
    |h_D| &\simeq \sqrt{0.7053\times10^{-3}} \simeq 0.0266\\
    |h_C| &\simeq \sqrt{0.1893\times10^{-3}} \simeq 0.0138\\
    \kappa &\simeq 1.363/2 \simeq 0.6815
\end{align}

\section{Weakly coupled chains}
\label{app:weakly_coupled}

\begin{figure}[h]
\begin{center}
\includegraphics[width=8cm]{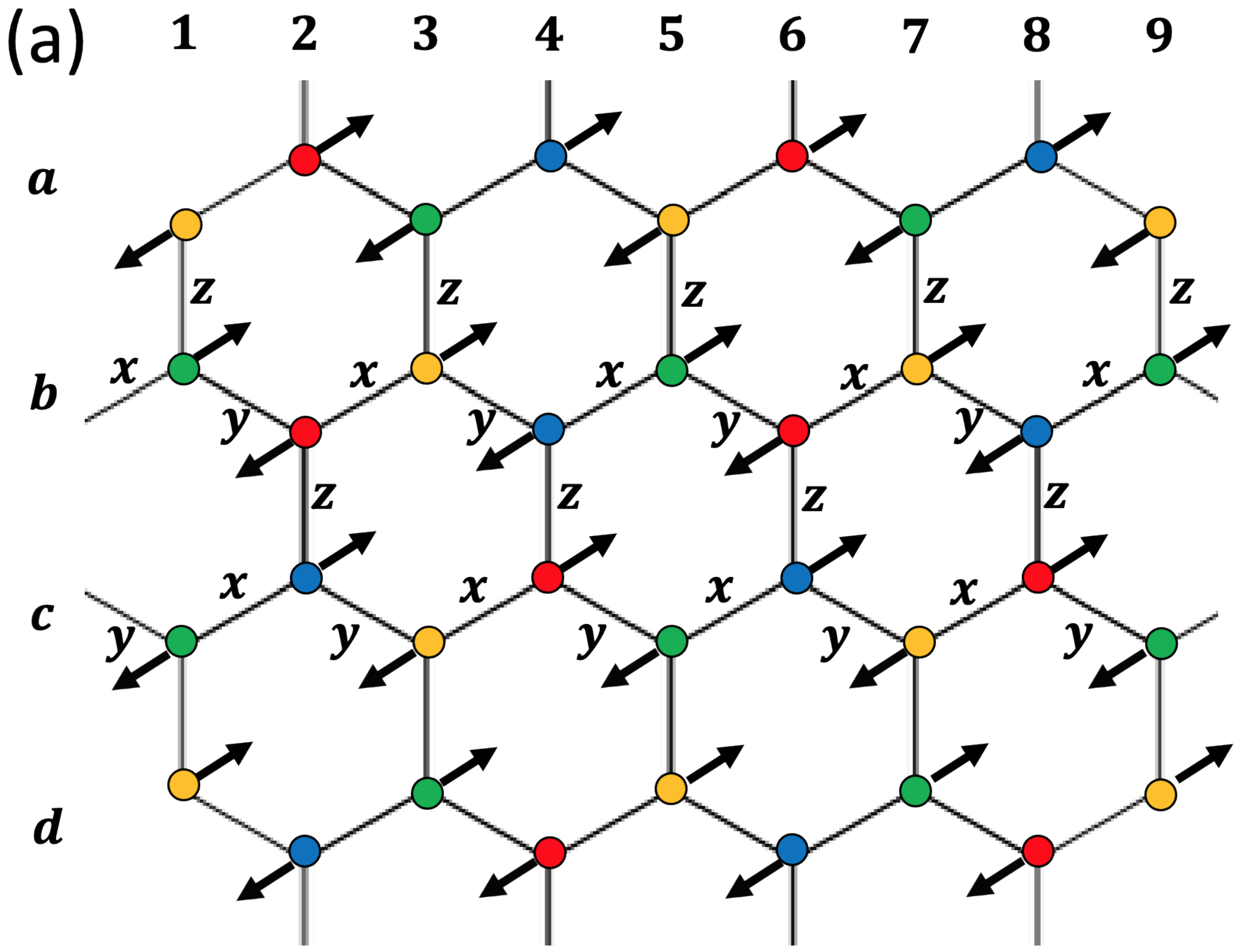}
\includegraphics[width=8cm]{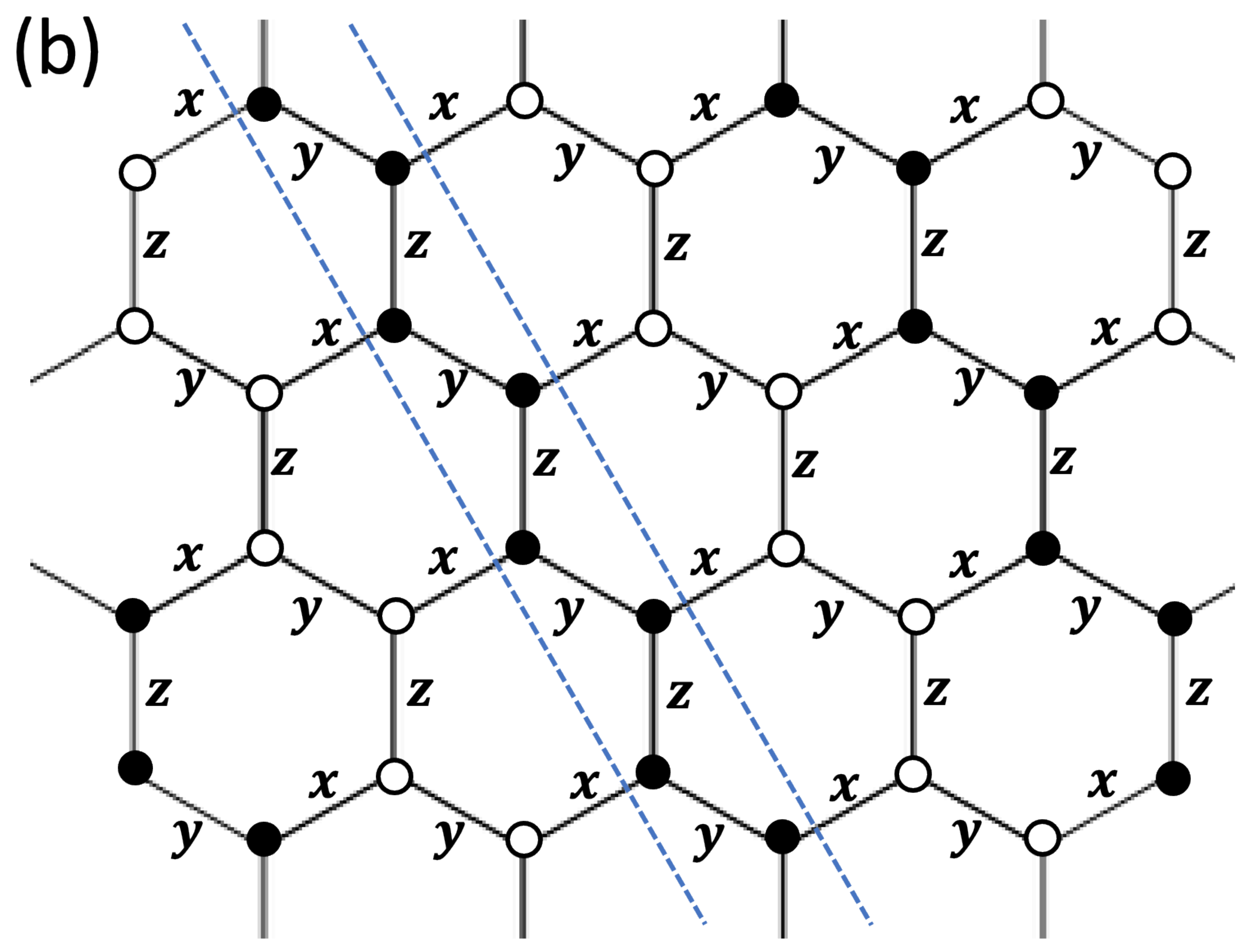}
\caption{(a)Four-sublattice rotation of the 2D Kitaev-Heisenberg-Gamma model on the honeycomb lattice, where the green, blue, yellow, and red circles represent the sites on which $R(\hat{y},\pi)$, $R(\hat{z},\pi)$, $R(\hat{x},\pi)$, and identity operation are applied, respectively;
(b) the zigzag order in the original frame. 
In (a), the black arrows denote the directions of the spin ordering with $\langle N^x\rangle\neq 0$ in the four-sublattice rotated frame;  the numbers ``1,2,3,..." and letters ``a,b,c,..." denote indices for the columns and rows, respectively,
where the rows are defined in the sense of the equivalent brick wall lattice. 
In (b), the solid and hollow circles denote spin magnetizations with  positive and negative components along $S^x$-axis in the spin space, respectively.  
} \label{fig:2D_zigzag}
\end{center}
\end{figure}

In this section, we discuss the recovery of zigzag phase in 2D Kitaev-Heisenberg-Gamma model by weakly coupling an infinite number of 1D chains.

\subsection{The mean field Hamiltonian}

Fig. \ref{fig:2D_zigzag} shows the four-sublattice rotation on the 2D honeycomb lattice, in which the green, blue, yellow, and red solid circles represent the sites on which $R(\hat{y},\pi)$, $R(\hat{z},\pi)$, $R(\hat{x},\pi)$, and identity operation are applied, respectively.
After the four-sublattice rotation, the interaction on bond $\gamma$ is 
\bea
H_{ij}=(K_\gamma+2J_\gamma)S_i^\gamma S_j^\gamma-J_\gamma \vec{S}_i\cdot \vec{S}_j +\epsilon(\gamma) \Gamma_\gamma (S_i^\alpha S_j^\beta+S_i^\beta S_j^\alpha),
\label{eq:2D_Hamiltonian}
\eea
in which $\epsilon(\gamma)=1$ when $\gamma=x,y,z$, and $\epsilon(\gamma)=-1$ when $\gamma=\bar{x},\bar{y},\bar{z}$;
$K_x=K_y=K$, $J_x=J_y=J$, $\Gamma_x=\Gamma_y=\Gamma$, $K_z=\alpha_0 K$, $J_z=\alpha_0 J$, $\Gamma_z=\alpha_0 \Gamma$.
When $\alpha_0\neq 1$, Eq. (\ref{eq:2D_Hamiltonian}) represents a dimerized Kitaev-Heisenberg-Gamma model. 
We will consider the limit $\alpha_0\ll 1$, so that the chains are weakly coupled. 

Let's consider row ``c", where the rows are defined in the sense of the equivalent brick wall lattice. 
When $\alpha_0=0$, the low energy theory of row c is described by the Luttinger liquid Hamiltonian $H_{LL,c}=\frac{v}{2} \int dx [\kappa^{-1} (\nabla \varphi)^2 +\kappa (\nabla \theta)^2]$.
Notice that  $N^x$ and $N^y$ have the smallest scaling dimension (both equal to $1/(2K)$), 
hence the leading instability in $H_{LL,c}$ is the Neel ordering in the $S^xS^y$-plane.
Since the chains are weakly coupled, 
a long range Neel order in row c will develop in the $S^xS^y$-plane.

In the low energy limit, the Luttinger liquid theory has an emergent U(1) symmetry, and different directions within the $S^xS^y$-plane are degenerate.
However, once a spin ordering is developed, there is an IR cutoff for the RG flow and the irrelevant couplings cannot be ignored since they no longer flow to zero.
The leading irrelevant term (i.e., the one having the smallest scaling dimension in the Luttinger liquid theory) which is consistent with all the symmetries in the symmetry group $G=\mathopen{<} T,R(\hat{y},\pi)T_a I,R(\hat{z},-\frac{\pi}{2})T_a \mathclose{>}$ can be checked to be $(N^+)^4+(N^-)^4\sim u\cos(4\sqrt{\pi}\theta)$, where $u$ is the coupling constant.  
The minima of this potential depend on the sign of $u$.
We note that the calculation of $u$ requires a high order perturbation treatment and we will not go into such difficult calculations.
At the moment, we will assume that $u<0$.
Then $u\cos(4\sqrt{\pi}\theta)$ is minimized when $\theta=\frac{\sqrt{\pi}}{2}n$, where $n\in\mathbb{Z}$.
Notice that when $n\in 2\mathbb{Z}$, we have $\langle N^x \rangle \neq 0$, $\langle N^y \rangle = 0$;
whereas when $n\in 2\mathbb{Z}+1$, we have $\langle N^x \rangle = 0$, $\langle N^y \rangle \neq 0$. 
There are four degenerate ground states, corresponding to $\langle N^x \rangle >0$, $\langle N^x \rangle < 0$, $\langle N^y \rangle >0$, and $\langle N^y \rangle <0$. 
Let's  take the ground state corresponding to $\langle N^x \rangle >0$, $\langle N^y \rangle =0$ as an illustrative example. 
Assuming a nonzero expectation value $\langle N^x \rangle$ in Eqs. (\ref{eq:bosonization_S1},\ref{eq:bosonization_S2},\ref{eq:bosonization_S3},\ref{eq:bosonization_S4}), we obtain
\begin{flalign}
&\langle \vec{S}_1 \rangle=\langle N^x \rangle(-a_C,-b_C,-c_C),~\langle \vec{S}_2 \rangle=\langle N^x \rangle(a_C,-b_C,-c_C),\nn\\
&\langle \vec{S}_3 \rangle=\langle N^x \rangle(-a_C,-b_C,c_C),~\langle \vec{S}_4 \rangle=\langle N^x \rangle(a_C,-b_C,c_C).
\label{eq:Nx_order_bosonize}
\end{flalign}

Before proceeding on, we make a comment on the unbroken symmetry group for the spin ordering in the 1D chain.
Since $T_{4a}$ is an unbroken symmetry and $|G/\mathopen{<}T_{4a}\mathclose{>}|=|D_{4d}|=16$, the unbroken symmetry group must be of order four. 
Consider the spin ordering pattern,
\begin{flalign}
&\langle \vec{S}_1 \rangle=\langle N^x \rangle\hat{x},~\langle \vec{S}_2 \rangle=-\langle N^x \rangle\hat{x},~\langle \vec{S}_3 \rangle=\langle N^x \rangle\hat{x},~\langle \vec{S}_4 \rangle=-\langle N^x \rangle\hat{x}.
\label{eq:order_Nx}
\end{flalign}
It can be checked the unbroken symmetry group $H$ of Eq. (\ref{eq:order_Nx}) is
\bea
H=\mathopen{<}T[R(\hat{z},-\frac{\pi}{2})T_a]^2,R(\hat{y},\pi)I\mathclose{>}.
\label{eq:unbroken_group}
\eea
However, Eq. (\ref{eq:order_Nx}) is not the only spin configurations consistent with Eq. (\ref{eq:unbroken_group}).
It can be worked out that the most general spin pattern having Eq. (\ref{eq:unbroken_group}) as the unbroken symmetry group is given by
\begin{flalign}
&\langle \vec{S}_1 \rangle=(-a,-b,-c),~\langle \vec{S}_2 \rangle=(a,-b,-c),\nn\\
&\langle \vec{S}_3 \rangle=(-a,-b,c),~\langle \vec{S}_4 \rangle=(a,-b,c).
\label{eq:order_Nx_general}
\end{flalign}
Comparing with Eq. (\ref{eq:Nx_order_bosonize}), it is clear that Eq. (\ref{eq:order_Nx_general}) has the same pattern as Eq. (\ref{eq:Nx_order_bosonize}).
However, we note that while the relation $\frac{a_C}{a}=\frac{b_C}{b}=\frac{c_C}{c}$ holds for small $\langle N^x \rangle$,
it is in general not true when the spin order is large,
since there can be other high order effects which renormalize the spin expectation values.

Next we perform a mean field analysis for the system of  weakly coupled 1D chains. 
For row c,  the mean field Hamiltonian is 
\bea
H_c=H_{cc}+H_{cb}+H_{cd},
\label{eq:H_c}
\eea
in which $H_{cc}$ is the intra-chain Hamiltonian which becomes the Luttinger liquid Hamiltonian at low energies
\bea
H_{cc}=\frac{v}{2} \int dx [\kappa^{-1} (\nabla \varphi)^2 +\kappa (\nabla \theta)^2],
\eea
and the interchain interactions are
\bea
H_{cb}&=&\sum_{n} \big[ \alpha_0 (K+J) S_{c,2+2n}^z\langle S_{b,2+2n}^z \rangle-\alpha_0 J (S_{c,2+2n}^x\langle S_{b,2+2n}^x \rangle+S_{c,2+2n}^y\langle S_{b,2+2n}^y \rangle)\nn\\
&&-\alpha_0 \Gamma (S_{c,2+2n}^x\langle S_{b,2+2n}^y \rangle+S_{c,2+2n}^y\langle S_{b,2+2n}^x \rangle)\big], 
\label{eq:Hcb}
\eea
and
\bea
H_{cd}&=&\sum_{n} \big[ \alpha_0 (K+J) S_{c,1+2n}^z\langle S_{d,1+2n}^z \rangle-\alpha_0 J (S_{c,1+2n}^x\langle S_{d,1+2n}^x \rangle+S_{c,1+2n}^y\langle S_{d,1+2n}^y \rangle)\nn\\
&&+\alpha_0 \Gamma (S_{c,1+2n}^x\langle S_{d,1+2n}^y \rangle+S_{c,1+2n}^y\langle S_{d,1+2n}^x \rangle)\big].
\label{eq:Hcd}
\eea
Notice that in Eqs. (\ref{eq:Hcb},\ref{eq:Hcd}), we have replaced the spin operators on rows b and d by their expectation values, in accordance with a mean field treatment. 
To proceed on, we note that the rule for bosonization formula is 
\begin{flalign}
&\text{Green: Eq. (\ref{eq:bosonization_S1})}, ~ \text{Blue: Eq. (\ref{eq:bosonization_S2})},\nn\\
&\text{Yellow: Eq. (\ref{eq:bosonization_S3})},~\text{Red: Eq. (\ref{eq:bosonization_S4})},
\end{flalign}
in which the colors denote the corresponding sublattices of the sites in Fig. \ref{fig:2D_zigzag}.
In addition, we assume $\langle N^x_b \rangle=-\langle N^x_c \rangle=\langle N^x_d \rangle$, because of the anti-ferromagnetic nature of the interchain coupling.
Then plugging in the bosonization formulas Eqs. (\ref{eq:bosonization_S1},\ref{eq:bosonization_S2},\ref{eq:bosonization_S3},\ref{eq:bosonization_S4}) and using $\langle N^x \rangle\neq 0$, Eqs. (\ref{eq:Hcb},\ref{eq:Hcd})  can be simplified.
The result is
\bea
H_{cb}+H_{cd}= -\frac{\lambda}{a} \langle N^x \rangle\int dx N^x,
\eea
in which $a$ is the lattice constant, and 
\bea
\lambda=- \alpha_0 [(K+J) c_C^2+J(a_C^2+b_C^2)-2\Gamma a_Cb_C]. 
\label{eq:lambda}
\eea
Notice that the coefficient $\lambda$ is positive since $J<0$ and $a_C\gg b_C,c_C$.

\subsection{Self-consistent solution}

The expectation value $\langle N^x \rangle$ can be solved in a self-consistent manner using the low energy mean field Hamiltonian
\bea
H_{MF}=\frac{v}{2} \int dx [\kappa^{-1} (\nabla \varphi)^2 +\kappa (\nabla \theta)^2]-\frac{\lambda}{a^3} \langle \cos(\sqrt{\pi}\theta) \rangle \int dx \cos(\sqrt{\pi}\theta),
\label{eq:H_MF}
\eea
in which $\lambda$ is given in Eq. (\ref{eq:lambda}), and $N^x=\cos(\sqrt{\pi}\theta)$ is used.

We use the variational method in Ref. \cite{Giamarchi2004_b} to solve the massive sine-Gordon model in Eq. (\ref{eq:H_MF}).
In the imaginary time path integral formalism, after integrating over $\varphi$, the action is
\bea
S=\frac{\kappa}{2} \int dxd\tau [\frac{1}{v} (\partial_\tau \theta)^2+v(\partial_x \theta)^2]-\frac{\lambda}{a^3} \langle \cos(\sqrt{\pi}\theta) \rangle \int dx \cos(\sqrt{\pi}\theta).
\eea
We first rewrite the action as
\bea
S=S_0+(S-S_0),
\eea
in which
\bea
S_0=\frac{\kappa}{2} \int dxd\tau [\frac{1}{v} (\partial_\tau \theta)^2+v(\partial_x \theta)^2+\frac{1}{v}\Delta^2\theta^2],
\eea
where $\Delta$ is the mass of the field. 
The partition function is
\bea
Z=\int D\theta e^{-S}=Z_0 \langle e^{-(S-S_0)} \rangle_0,
\eea
in which $Z_0=\int D\theta e^{-S_0}$ and $\langle ... \rangle_0=\frac{1}{Z_0}\int D\theta e^{-S_0} (...)$.
In the variational approach,  the following inequality is used
\bea
F \leq  F^\prime = F_0 +\frac{1}{\beta} \langle S-S_0\rangle_0,
\eea
in which $F=\ln Z$, $F^\prime$, $F_0=\ln Z_0$ are the free energy, the variational free energy, and the free energy for the action $S_0$, respectively,
and $\beta$ is the temperature. 

Denote $G(\vec{q})$ as
\bea
G(\vec{q})=\frac{\kappa^{-1}}{\frac{1}{v}\omega_n^2+vk^2+\frac{1}{v}\Delta^2},
\eea
in which $\vec{q}=(\omega_n,k)$ where $\omega_n$ is the Matsubara frequency, and $k$ is the 1D wavevector.
Then we have
\bea
F^\prime=-\frac{1}{\beta} \sum_{\vec{q},k>0} \log [G(\vec{q})]+\frac{\kappa}{2}\frac{1}{\beta} \sum_{\vec{q}} (\frac{1}{v}\omega_n^2+vk^2) G(\vec{q})-\frac{1}{\beta}\frac{\lambda}{a^3} \langle \cos(\sqrt{\pi}\theta) \rangle \beta L e^{-\frac{\pi}{2\beta L}\sum_{\vec{q}} G(\vec{q})},
\eea
in which $L$ is the length of the system.
The optimal $\Delta$ can be determined by the saddle point equation
\bea
\frac{\partial F^\prime}{\partial G(\vec{q})}=0,
\eea
which gives 
\bea
G^{-1}(\vec{q})=\kappa (\frac{1}{v}\omega_n^2+vk^2+\frac{\Delta^2}{v}),
\eea
where
\bea
\frac{\kappa\Delta^2}{v}=\frac{\pi\lambda}{a^3} \langle \cos(\sqrt{\pi}\theta) \rangle e^{-\frac{\pi}{2\beta L}\sum_{\vec{q}} \frac{v\kappa^{-1}}{\omega_n^2+v^2k^2+\Delta^2} }.
\label{eq:self_consistent_eq}
\eea
In what follows, we will consider the zero temperature case. 
Denoting $\Lambda$ to be the UV cutoff in the theory (which is on the order of $1/a$), 
and assuming $\Delta\ll \Lambda$,
we have
\bea
\frac{\pi}{2\beta L}\sum_{\vec{q}} \frac{v\kappa^{-1}}{\omega_n^2+v^2k^2+\Delta^2}&=&\frac{\pi}{2(2\pi)^2} \int d\vec{q} \frac{v\kappa^{-1}}{\omega_n^2+v^2k^2+\Delta^2}\nn\\
&\simeq& (4\kappa)^{-1} \ln [v\Lambda/\Delta].
\eea
Thus Eq. (\ref{eq:self_consistent_eq}) becomes
\bea
\Delta^2=\frac{\pi v\lambda}{\kappa a^3} \langle \cos(\sqrt{\pi}\theta) \rangle \big(\frac{\Delta}{v\Lambda}\big)^{(4\kappa)^{-1}},
\eea
which yields
\bea
\Delta = v\Lambda \big[\frac{\pi\lambda\langle \cos(\sqrt{\pi}\theta) \rangle}{v\kappa \Lambda^2a^3}\big]^{\frac{1}{2-(4\kappa)^{-1}}}.
\eea

On the other hand, using the action $S_0$, the expectation value of $\cos(\sqrt{\pi}\theta)$ can be calculated as
\bea
\langle \cos(\sqrt{\pi}\theta) \rangle=e^{-\frac{\pi}{2\beta L}\sum_{\vec{q}} G(\vec{q})}=\big(\frac{\Delta}{v\Lambda}\big)^{(4\kappa)^{-1}}.
\eea
Hence self-consistency requires
\bea
\langle \cos(\sqrt{\pi}\theta) \rangle
=\big[\frac{\pi\lambda\langle \cos(\sqrt{\pi}\theta) \rangle}{v\kappa \Lambda^2a^3}\big]^{\frac{(4\kappa)^{-1}}{2-(4\kappa)^{-1}}}
\eea
from which $\langle \cos(\sqrt{\pi}\theta) \rangle$ can be solved as
\bea
\langle \cos(\sqrt{\pi}\theta) \rangle=\big[\frac{\pi\lambda}{v\kappa \Lambda^2a^3}\big]^{\frac{1}{8\kappa-2}}\sim (\alpha_0)^{\frac{1}{8\kappa-2}}.
\label{eq:sol_Nx}
\eea

\subsection{The 2D zigzag order}

The  pattern of the spin ordering on the 2D lattice can be written in a concise expression.
For this, it is useful to note the following relations:
\bea
\sqrt{2}\cos(\frac{\pi}{2}(j+\frac{1}{2}))&:& ( -1, -1, 1, 1)\nn\\
\sqrt{2}\sin(\frac{\pi}{2}(j+\frac{1}{2}))&:&  (1, -1, -1 , 1),
\eea
in which the components of the vectors on the right hand side corresponds to the value of the functions at $j=1$, $j=2$, $j=3$, $j=4$, respectively.
Denote row c in Fig. \ref{fig:2D_zigzag} to be row $1$, and row index raises moving upward. 
Then within the four-sublattice rotated frame, the spin pattern is given by
\bea
\langle \vec{S}_{2m+1,n} \rangle = \langle N^x \rangle\left(\begin{array}{c}
(-)^n a_C\\
-b_C\\
\sqrt{2}\cos(\frac{\pi}{2}(n-2m+\frac{5}{2})) c_C
\end{array}\right),~
\langle \vec{S}_{2m,n} \rangle = \langle N^x \rangle\left(\begin{array}{c}
(-)^{n+1} a_C\\
b_C\\
-\sqrt{2}\sin(\frac{\pi}{2}(n-2m+\frac{1}{2})) c_C
\end{array}\right). 
\label{eq:spin_order_4rotation}
\eea
When $K^\prime$ and $\Gamma$ are small compared with $|J|$,
since $a_C$ dominates over $b_C$ and $c_C$, $\langle\vec{S}_{(m,n)}\rangle$ is mainly along $S^x$-direction, with minor components along $S^y$- and $S^z$-directions.  
The black arrows in Fig. \ref{fig:2D_zigzag} (a) shows the $S^x$-components for the expectation values of the local spin operators on the honeycomb lattice. 

The spin pattern in Eq. (\ref{eq:spin_order_4rotation}) can be transformed back to the original frame without four-sublattice rotation.
The result is
\bea
\langle \vec{S}_{2m+1,n} \rangle = \sqrt{2}\sin(\frac{\pi}{2}(n-2m+\frac{1}{2}))\langle N^x \rangle\left(\begin{array}{c}
a_C\\
-b_C\\
c_C
\end{array}\right),~
\langle \vec{S}_{2m,n} \rangle = \sqrt{2}\cos(\frac{\pi}{2}(n-2m+\frac{1}{2})) \langle N^x \rangle\left(\begin{array}{c}
 a_C\\
-b_C\\
 c_C
\end{array}\right). 
\label{eq:spin_order_4rotation}
\eea
Alternatively, Eq. (\ref{eq:spin_order_4rotation}) can be concisely written as
\bea
\vec{S}_{m,n}=\sqrt{2}\cos(\frac{\pi}{2}(n-m)+\frac{1}{2})\langle N^x \rangle (a_C,-b_C,c_C).
\eea
The solid and hollow circles in Fig. \ref{fig:2D_zigzag} (b) denote the local spin expectation values which have positive and negative components along $S^x$-direction, respectively.
Clearly, the zigzag chain sandwiched by the two dashed blue lines in Fig. \ref{fig:2D_zigzag} (b) is FM along $S^x$-axis in the spin space.
Notice that Fig. \ref{fig:2D_zigzag} (b) is exactly the zigzag order on the 2D honeycomb lattice.
